\documentclass[prb,a4paper,twocolumn,floatfix]{revtex4-1}
\usepackage{color}
\usepackage[usenames,dvipsnames]{xcolor}
\usepackage{graphicx}
\usepackage{latexsym}
\usepackage[]{amsmath}
\usepackage{epstopdf}
\usepackage{amssymb}
\usepackage{array}

\usepackage{dcolumn}
\usepackage{bm}

\begin{document}

\title{Combining micro- and macroscopic probes to untangle single-ion and spatial exchange anisotropies in a $S = 1$ quantum antiferromagnet}

\author{ Jamie Brambleby$^{1\ddag}$, Jamie L. Manson$^{2,3*}$, Paul A. Goddard$^1$,
Matthew B. Stone$^4$, Roger D. Johnson$^{5,6}$,
Pascal Manuel$^6$,  Jacqueline A. Villa$^2$, Craig M. Brown$^3$,
Helen Lu$^7$, Shalinee Chikara$^7$,
Vivien Zapf$^7$, Saul H. Lapidus$^8$, Rebecca Scatena$^9$, 
Piero Macchi$^9$, Yu-sheng Chen$^{10}$, Lai-Chin Wu$^{10}$ 
and John Singleton$^{7,\aleph}$}

\affiliation{$^1$Department of Physics, University of Warwick, Gibbet Hill Road, 
Coventry CV4 7AL, United Kingdom\\
$^2$Department of Chemistry and Biochemistry, 
Eastern Washington University, Cheney, WA 99004, U.S.A.\\
$^3$NIST Center for Neutron Research, 
National Institute of Standards and Technology, 
Gaithersburg, MD 20899, U.S.A.\\
$^4$Quantum Condensed Matter Division, Oak Ridge National Laboratory, 
Oak Ridge, TN 37831, U.S.A.\\
$^5$University of Oxford, Department of Physics,
The Clarendon Laboratory, Parks Road, Oxford, OX1~3PU,
United Kingdom\\
$^6$ISIS Pulsed Neutron Source, STFC Rutherford Appleton Laboratory, 
Didcot, Oxfordshire OX11 0QX, United Kingdom\\
$^7$National High Magnetic Field Laboratory, MS-E536, 
Los Alamos National Laboratory, Los Alamos, NM~87545, U.S.A.\\
$^8$X-ray Sciences Division, Advanced Photon Source, 
Argonne National Laboratory, Argonne, IL 60439, U.S.A.\\
$^9$Department of Chemistry and Biochemistry, University of Bern, 
3012 Bern, Switzerland\\ 
$^{10}$ChemMatCARS, Advanced Photon Source, 
Argonne National Laboratory, Argonne, IL 60439, U.S.A.}

\begin{abstract}
The magnetic ground state of the quasi-one-dimensional 
spin-1 antiferromagnetic chain is sensitive to the relative 
sizes of the single-ion anisotropy ($D$) and the intrachain 
($J$) and interchain ($J'$) exchange interactions. 
The ratios $D/J$ and $J'/J$ dictate the material's placement 
in one or other of three competing phases: a Haldane gapped phase, 
a quantum paramagnet and an XY-ordered state, with a 
quantum critical point at their junction. 
We have identified [Ni(HF)$_2$(pyz)$_2]$SbF$_6$, 
where pyz = pyrazine, as a candidate in which this behavior 
can be explored in detail. Combining neutron scattering 
(elastic and inelastic) in applied magnetic fields of up to 10~tesla 
and magnetization measurements in fields of up to 60~tesla 
with numerical modeling of experimental observables, 
we are able to obtain accurate values of all of the parameters 
of the Hamiltonian [$D = 13.3(1)$~K, $J = 10.4(3)$~K and 
$J' = 1.4(2)$~K], despite the polycrystalline nature of the sample. 
Density-functional theory calculations result in similar couplings 
($J = 9.2$~K, $J' = 1.8$~K) and predict that the majority of the total 
spin population of resides on the Ni(II) ion, while the remaining 
spin density is delocalized over both ligand types. The general 
procedures outlined in this paper permit phase boundaries and 
quantum-critical points to be explored in anisotropic systems 
for which single crystals are as yet unavailable. 
\end{abstract}


\maketitle

\section{Introduction}
The possibility of arranging interacting magnetic moments in 
chains or planes has excited theorists and experimentalists 
alike for many years. The pioneering work by Kosterlitz and 
Thouless on $XY$-like spins in two dimensions~\cite{a1,b1}, 
and by Haldane on integer-spin chains~\cite{a2,b2} has had 
far-reaching implications, culminating in the award of the 
2016 Nobel Prize in Physics~\cite{nobel}. More recently, 
interest in predicting and controlling the magnetic ground state of 
$S = 1$ quantum magnets has been fueled by the realization 
of a myriad of magnetic phases in a series of metal-organic 
coordination compounds. This includes the observation of 
field-induced Bose-Einstein condensation in 
NiCl$_2$-4SC(NH$_2$)$_2$~[\onlinecite{a3,a4,a5}], as well 
as the development of a Haldane phase in both [Ni(C$_2$H$_8$N$_2$)$_2$NO$_2$]ClO$_4$~[\onlinecite{a6}] 
and [Ni(HF$_2$)(3-Clpy)$_4$]BF$_4$ (Clpy = C$_5$H$_4$NCl = chloropyridine)~[\onlinecite{a7,a8,a9}]. Ground-state 
diversity is attributable to the interplay between the 
single-ion anisotropy $(D)$ and Heisenberg spin-exchange 
interactions $(J)$ in these materials, which are both determined 
(in part) by the lattice geometry~\cite{a9}. The flexibility
offered by the crystal structures of quasi-one dimensional coordination 
polymers renders them ideal systems to advance our understanding 
of the quantum-critical phenomena associated with $S = 1$ systems.

The magnetic ground state of a quasi-one-dimensional (Q1D) $S = 1$ 
antiferromagnet (AFM) is particularly sensitive to both the precise nature 
of $D$ and its strength compared to that of the intrachain 
Heisenberg spin-exchange interaction~\cite{a7}. The magnetic 
properties in a magnetic field $(B\approx \mu_0H)$ 
are summarized by the Hamiltonian 
\[
{\cal H} =\sum_{<i,j>} J{\bf S}_i.{\bf S}_j
+\sum_{<i,j'>} J'{\bf S}_i.{\bf S}_{j'} 
\]
\begin{equation}
+
\sum_i \left[D(S_i^z)^2 + g\mu_{\rm B}{\bf S}_i.{\bf B}\right],
\label{eqn1}
\end{equation}
where {\bf S} is the spin of each ion $(i)$, $\langle i,j\rangle$ denotes a sum 
over nearest neighbors, $J'$ is the strength of the interchain interaction, 
a primed index in the summation describes the interaction with a nearest 
neighbor in an adjacent chain and $g$ is the isotropic $g$-factor.

Unlike classical systems, ideal $S = 1$ chains $(J' = 0)$ 
are vulnerable to strong quantum fluctuations~\cite{a8}, which 
can have a profound influence on the magnetic ground state and act 
to suppress long-range order. Quantum Monte-Carlo (QMC) simulations 
predict that for easy-plane anisotropy a Haldane ground state 
gives way to quantum paramagnetism (QP) as the $D/J$ ratio increases, 
while the effects of $J'$ are to alleviate the quantum disorder and induce 
an $XY$-AFM (or N\'{e}el if $D = 0$) ordered phase~\cite{a7}. 
The three magnetic phases are expected to converge at a 
quantum critical point (QCP)~[\onlinecite{a10}], located at 
$D/J = 0.97$ for purely 1D chains~\cite{a9}. Therefore, the ability to 
measure $J$, $J'$ and $D$ precisely, in addition to obtaining 
an unambiguous experimental determination of the magnetic 
ground state in real systems, is a crucial step towards testing the 
theoretical predictions of quantum phenomena in Q1D $S = 1$ chains. 
However, as we now describe, this has proved challenging in the case 
of polycrystalline samples, particularly for those systems in which 
the exchange and anisotropy energies are similar in magnitude. 
It can often take time to hone synthetic methods sufficiently 
to obtain single crystals large enough for many measurement 
techniques. Therefore it is frequently the newest, most exciting families 
of materials that are most difficult to characterize.

In the absence of a magnetic field and strong spin-exchange interactions, 
systems described by Eqn.~\ref{eqn1} are dominated by single-ion anisotropy. 
This energy term acts to remove the spin-microstate degeneracy of 
paramagnetic Ni(II) ions $(m_s = 0, \pm1)$ and is dependent upon 
both the metal-ligand electronic structure and the spin-density 
distribution. For hexa-coordinated Ni(II) coordination complexes, 
this zero-field splitting (ZFS) of energy levels can result in a singlet 
$(D > 0)$ or doublet ground state with $D$-values that have been 
found to span the range $-32 \leq D \leq 20$~K~[\onlinecite{a11,a12}]. 
So long as the Ni$\cdots$Ni spin-exchange interactions are weak, 
the $D$-value in complexes of this type may be characterized by 
magnetic susceptibility, magnetization, and heat capacity 
measurements. However, a reliable estimation of both 
the size {\it and} sign of $D$ in polycrystalline samples is only 
possible via these techniques if one can apply magnetic fields 
of a sufficient strength to significantly align the spins~\cite{will}. 
Electron-spin resonance (ESR) is also frequently used to 
determine the anisotropy and can work well for powdered 
samples, but only provided the frequency-field combination 
that matches the ZFS can be achieved.

For exchange-coupled systems, the sensitivity of bulk 
thermodynamic probes to the spin correlations further 
complicates the extraction of a unique value for the 
single-ion anisotropy. A resolution to this problem 
is offered by a microscopic probe, such as inelastic 
neutron scattering (INS), which is well suited to 
differentiating the effects of spin-exchange 
interactions from those of single-ion anisotropy. 
The origin of spin excitations (spin-wave or 
crystal-field levels) may be discriminated 
by their wave-vector $({\bf Q})$ and energy-transfer 
$(E)$ dependence~\cite{a13}. 
In the past, successful treatment 
of INS data has relied on instances 
of $D$-only models (ignoring $J$), as found 
in high-nuclearity Mn(III) complexes\cite{a14,b14}, 
$J$-only models (ignoring $D$) in some Co(II) 
complexes~\cite{a15,b15,c15} or $J > D$ as 
determined in Q1D MnCl$_2$(urea)$_2$~[\onlinecite{a16}]. 
To the best of our knowledge, analysis of powder INS 
spectra has not been successfully tested in multi-parameter 
systems where $D\approx J$ until now. Without INS data, 
density-functional theory (DFT) is often implemented to 
validate thermodynamic parameters according to a 
prescribed Hamiltonian (such as Eqn.~\ref{eqn1}). 
The results, however, are sensitive to the basis set 
employed and require experimental support if one is 
to have confidence in the outcome~\cite{a17,b17,c17}.
 
Here we describe a complete experimental procedure to 
determine the $H,T$ phase diagram and
all of the parameters of the spin Hamiltonian
of [Ni(HF$_2$)(pyz)$_2$]SbF$_6$, despite the lack of suitable 
single crystals. This material is composed of linear HF$_2^-$ 
pillars that mediate an intrachain Ni---Ni exchange coupling 
$(J)$, while bridging pyrazine ligands provide four equivalent 
interactions $(J')$ to neighboring chains~\cite{a18}. 
The material enters an AFM ordered phase below 12.2~K 
and exhibits $D/J \approx 1$ along with a strong spatial 
exchange anisotropy (predicted from DFT to be 
$J'/J \approx 0.1$)~[\onlinecite{a18}]. It therefore 
provides a rare opportunity to study the magnetic 
properties of a system close to the three competing Q1D ground states. 

Below the ordering temperature, elastic neutron scattering 
of [Ni(HF$_2$)(pyz)$_2$]SbF$_6$ reveals that the zero-field 
magnetic structure is that of a 3D $XY$-AFM ground state. 
Based on this result, the anisotropic critical field observed in 
powder magnetization measurements can then be interpreted 
within an easy-plane, mean-field picture to initially estimate 
values of $D = 13.1(3)$~K and $n\langle J \rangle = 22.1(2)$~K ($n$ = number 
of magnetic nearest neighbors and $\langle J \rangle$ = average spin-exchange 
interaction strength). Applying Eqn.~\ref{eqn1} in SPINW~\cite{a35}, 
we model powder INS spin-wave spectra to deconvolute the 
two distinct AFM contributions to 
$n\langle J \rangle$ to yield $J = 10.4(3)$~K and $J' = 1.4(2)$~K that we 
assign to Ni---FHF---Ni and Ni---pyz---Ni interactions, respectively. 
Given these parameters, we find good agreement with the 
predictions of QMC calculations for the low-temperature phase 
and this result is used to explain the form of the field-temperature 
phase diagram revealed by heat capacity and magnetization 
measurements. We compare these parameters to those 
obtained from low-field magnetic susceptibility 
measurements and demonstrate the shortcomings
 in modeling these data in the absence of other information. 
Lastly, we provide a detailed analysis of the spin-density 
distribution and exchange-coupling constants as 
predicted by periodic DFT calculations. 

\section{Experimental methods}
\subsection{Design strategy}
The tunability offered by metal-organic systems renders them ideal testbeds 
for which to investigate numerous quantum theories relevant to 
reduced-dimensionality physics. 
Our current endeavors are focused on {\it M}(II)-based 
[{\it M} = Co $(S = \frac{3}{2})$, Ni $(S = 1)$, Cu $(S = \frac{1}{2})$] 
coordination polymers, self-assembled from strong 
charge-assisted H-bonds 
({\it e.g.}, F$\cdots$H$\cdots$F)~[\onlinecite{a7,a8,a18,a19,a20,a21,a22,a23,a24,a25,a26,a27}]. Weaker O-H$\cdots$F types 
have also been examined~\cite{a28,a29,a30,a31,a32,a33,a34}. 
The use of crystal engineering to manage 
such interactions is a promising 
path forward in the experimental search for exotic phases of quantum matter. 
Among the various Ni(II) systems reported thus far, 
[Ni(HF$_2$)(pyz)$_2$]SbF$_6$ (pyz = pyrazine)~\cite{a18},
$\alpha -$ and $\beta -$polymorphs of 
[Ni(HF$_2$)(pyz)$_2$]PF$_6$ and 
[Ni(HF$_2$)(3-Clpy)$_4$]BF$_4$ have been synthesized and 
characterized~\cite{a7,a8,a18,a19}. 
The latter compound forms isolated Q1D Ni---FHF---Ni chains with 
bent and asymmetric HF$_2^-$ bridges~\cite{a7,a8}. 
Moreover, the suppressed long-range magnetic order (LRO) 
and $D/J \approx 1$ are unique to this material, as it is the 
only Ni(II)-chain proximate to the $D/J = 0.97$ QCP, 
which separates Haldane, $XY$-AFM, and QP phases~\cite{a9}.

In [Ni(HF$_2$)(pyz)$_2$]$X$, the presence of additional, albeit weak, 
interchain Ni---pyz---Ni exchange pathways 
promote LRO~\cite{a18,a19} 
below temperatures $\approx 6$~K ($X = \alpha -$PF$_6$), 7~K 
($X = \beta -$PF$_6$), and 12.2~K ($X  = $ SbF$_6$). 
For each compound, a substantial zero-field splitting (ZFS) 
is also anticipated due to the 
existing NiN$_4$F$_2$ core; {\it e.g.}, an easy-plane $D = 20$~K 
has been determined for Q1D Ni(SiF$_6$)(vinim)$_4$ 
(vinim = 1-vinylimidazole)~\cite{a12}. 
\subsection{Synthesis}
All reagents were obtained from commercial suppliers and used as received. 
Plasticware was used throughout the entire 
synthetic process~\cite{HCl}. 
In a typical synthesis, NiF$_2$ (1.775~mmol, 302.0~mg) 
was dissolved separately in 3~mL of aqueous-HF (48-51~\% by weight)
while NH$_4$HF$_2$ (1.775~mmol, 106.5~mg), LiSbF$_6$ 
(1.775~mmol, 431.8~mg), and pyrazine (3.550~mmol, 283.7~mg) 
were dissolved together in a separate beaker containing
 2~mL of aqueous HF and 1~mL H$_2$O. 
The NiF$_2$ solution was slowly mixed with the ligands 
to give a green solution that was covered with a perforated 
wax film to allow slow evaporation of the solvent. 
On standing at room temperature for 3~weeks, 
a blue powder formed on the bottom and walls 
of the beaker. The solid was collected by vacuum 
filtration, washed with 2~mL of H$_2$O, 2~mL of 
ethanol, and 2~mL of diethyl ether to assist drying. 
A pale blue powder, consisting of microcrystalline plates 
($10 \times 10 \times 2~\mu$m$^3$ in average size), 
was obtained in high yield ($> 80$~\%) based on Ni(II) content). 
Further details are available in the Supplementary Information~\cite{SI}.
While these microcrystals were suitable for the 
synchrotron X-ray structural study they were 
much too small for single-crystal thermodynamic 
and neutron-scattering measurements.
 
For the neutron-scattering experiments, 
synthesis of a partially deuterated phase of 
[Ni(HF$_2$)(pyz-d$_4$)$_2$]SbF$_6$, was necessary to reduce the large 
incoherent scattering cross-section due to protons ({\it i.e.} the $^1$H
nuclei). The synthesis was carried out as described above, 
except pyz was replaced by pyz-d$_4$. To produce a 
1.8~g sample, suitable amounts of HF(aq) and other reagents 
were used accordingly. X-ray powder diffraction patterns and 
magnetic susceptibility data were found to be very similar for the 
hydrogenated and deuterated phases. 

\subsection{Microcrystal X-ray diffraction}
Experiments were conducted on the ChemMatCARS 15-ID-B 
beamline of the Advanced Photon Source (APS) at 
Argonne National Laboratory (ANL). 
A microcrystal of [Ni(HF$_2$)(pyz)$_2$]SbF$_6$ measuring 
$10\times 10\times 2~\mu$m$^3$ was selected from a bulk 
sample using a cryo-loop and mounted on a Bruker D8 fixed-chi 
X-ray diffractometer equipped with an APEX II CCD area detector. 
The sample was cooled to 15(2)~K using a He-cryojet. 
Synchrotron radiation with a beam energy of 32.2~keV 
($\lambda = 0.38745$~\AA) was used, and the beam size 
at the sample  was $0.1 \times 0.1$~mm$^2$. 
The distance between sample and detector was set at 60~mm. 
A total of 720 frames were collected at $\theta = -5^{\circ}, 140^{\circ}$ 
and $180^{\circ}$ with the $\phi$-angle scanned over $180^{\circ}$ 
at intervals of $0.5^{\circ}$. Data collection and integration 
were performed using the APEX II software suite. 
Data reduction employed SAINT~\cite{a36}. 
Resulting intensities were corrected for absorption by 
Gaussian integration (SADABS)~\cite{a37}. 
The structural solution (XT)~[\onlinecite{a38}] and 
refinement (XL)~[\onlinecite{a39}] were carried out with 
SHELX software using the XPREP utility for the space-group determination. 
Considering systematic absences, the crystal structure was 
solved in the tetragonal space group $P4/nmm$ 
(\#129, origin choice 2)~[\onlinecite{a40}]. 
Pyrazine H-atoms were placed in idealized positions 
and allowed to ride on the carbon atom to which they are attached. 
All non-hydrogen atoms were refined with 
anisotropic thermal displacement parameters. 

\subsection{Elastic neutron scattering}
 Magnetic diffraction patterns were recorded on the 
WISH diffractometer (ISIS, Rutherford Appleton Laboratory, UK)~\cite{a41}.
As mentioned above, a partially deuterated sample, 
[Ni(HF$_2$)(pyz-d$_4$)$_2$]SbF$_6$, of mass 1.8~g was loaded into a 
cylindrical vanadium can and placed in an Oxford Instruments cryostat 
with a base temperature of 1.5~K. Diffraction data were 
collected over the temperature interval $1.5-20$~K, with long counting 
times (8~hrs) at 1.5~K and 20~K. Intermediate temperature points 
were measured with an exposure time of 2~hrs. 
Rietveld refinements were performed using FULLPROF~\cite{a42}. 
All atoms were refined using isotropic thermal displacement parameters. 

\subsection{Heat capacity}
Heat capacity $(C_p)$ measurements were carried out using a 9~T 
Quantum Design PPMS, with a 1.91(5)~mg powder sample of 
[Ni(HF$_2$)(pyz)$_2$]SbF$_6$ that was pressed into a pellet, 
secured to a sapphire stage with Apiezon-N grease and held in 
contact with a large thermal bath. 
Measurements of $C_p$ were performed using the traditional 
relaxation method~\cite{a43}. For this technique, a heat 
pulse ($\approx 1~\%$ of the thermal bath temperature) was 
applied to the stage and $C_p$ evaluated by measuring the time 
constant of the thermal decay curve. The 
heat capacities of the Apiezon-N grease and sample platform were 
measured separately and subtracted from the total 
to obtain the heat capacity of the sample. 

\subsection{Temperature-dependent linear susceptibility}
Linear susceptibility ($M/\mu_0 H$, where $M$ is the magnetization) 
measurements were made for temperatures in the range 
$1.9 \leq T \leq 50$~K and fields $\mu_0 H \leq 13$~T 
using a Quantum Design Physical Property Measurement System (PPMS) 
equipped with a vibrating sample magnetometer (VSM). 
A 2~mg powder sample was loaded into a brass holder, 
and mounted in the VSM transport. The sample was initially cooled 
in zero applied field to $T= 1.9$~K. 
Data were collected upon warming and corrected for diamagnetic 
contributions from the sample and background. 

\subsection{Pulsed-field magnetization}
Measurements of the powder magnetization of 
[Ni(HF$_2$)(pyz)$_2$]SbF$_6$ up to 60~T made use of a 
1.5~mm bore, 1.5~mm long, 1500-turn compensated-coil 
susceptometer, constructed from a 50~gauge high-purity 
copper wire~\cite{a23}. When the sample is within the coil, 
the signal voltage $V$ is proportional to ${\rm d}M/{\rm d}t$, 
where $t$ is time. Numerical integration of $V$ is used to evaluate $M$. 
The sample is mounted within a 1.3~mm diameter ampoule 
that can be moved in and out of the coil. 
Accurate values of $M$ are obtained by subtracting empty-coil data 
from that measured under identical conditions with the sample present. 
The susceptometer was placed inside a $^3$He cryostat providing a base 
temperature of 0.5~K. The magnetic field was measured by integrating 
the voltage induced in a 10-turn coil calibrated by observing the 
de Haas-van Alphen oscillations of the belly orbits of the copper 
contained in the susceptometer coil~\cite{a23}.
 
\subsection{Inelastic neutron scattering}
Two INS measurements were performed using the disk-chopper 
spectrometer (DCS) located at the NIST Center for 
Neutron Research (NCNR)~\cite{a44}. 
Both measurements used neutrons of wavelength 3.7~\AA 
~to scatter 
from a 1.8~g, partially deuterated powder sample of 
[Ni(HF$_2$)(pyz-d$_4$)$_2$]SbF$_6$ loaded in an aluminum 
sample can sealed under a helium atmosphere. 
The first measurement was conducted in zero field 
to probe the temperature-dependence of INS spectra at
$T = 1.6$~K, 10~K, and 20~K. 
The magnetic spin excitations were examined by subtracting the 20~K 
({\it i.e.} above the ordering temperature) data from the 1.6~K measurement. 

\begin{figure}[t]
\centering
\includegraphics[width=8.5cm]{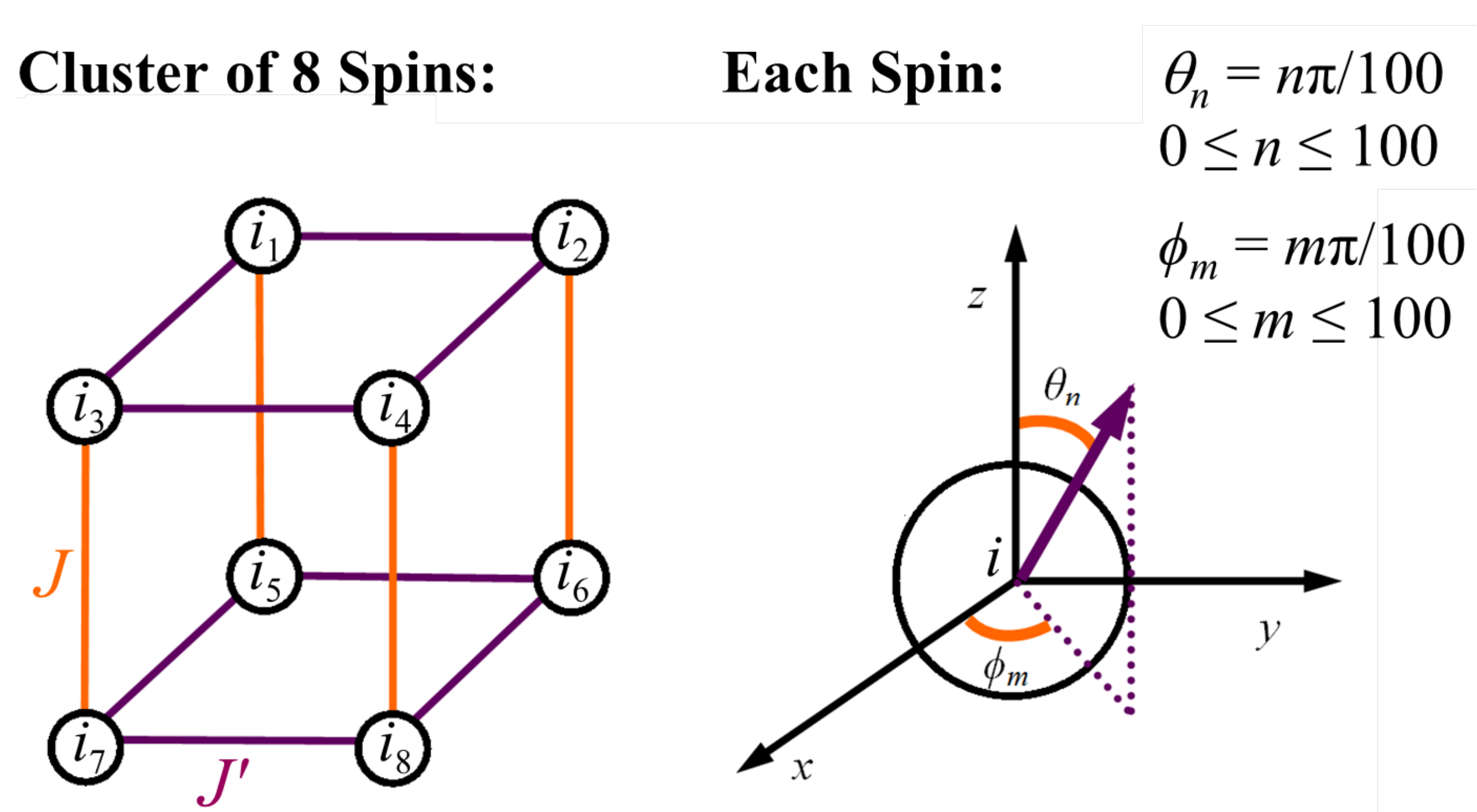}
\caption{The $2\times 2$ planar arrangement of Ni(II) ions used 
in the simulation of the magnetization. Each ion $(i)$ is represented 
by a vector where the two degrees of freedom are the polar angles 
$\theta_n$ and $\phi_m$, each limited to 100 evenly spaced values.
$J$ and $J'$ correspond to Ni---FHF---Ni and Ni---pyz---Ni exchange 
interactions, respectively. 
}
\label{fig1}
\end{figure}

Thermodynamic parameters were determined by comparing 
the measured powder spin-wave spectra against a series of 
spectra simulated using the SPINW analysis package~\cite{a35}, 
assuming the $XY$ magnetic structure determined by elastic neutron 
scattering. The second measurement surveyed the magnetic-field 
dependence of spin excitations at $T = 1.6$~K and 20~K in 
magnetic fields of $\mu_0 H = 0, 3, 6,$ and 10~T. To account 
for the time-independent background of the measurements, 
an overall value of 27 counts per hour per detector was 
subtracted from the data. This value was not subtracted in 
cases where data are plotted as the difference between 
two temperatures or two fields.

\section{Theory, calculations and simulations}
\subsection{Monte-Carlo simulation of high-field magnetization}
We compute the powder-average magnetization of 
[Ni(HF$_2$)(pyz)$_2$]SbF$_6$ using a classical Monte-Carlo 
routine written in MATLAB~\cite{a45}. 
A cluster of 8 ions is arranged in two sets of $2\times 2$ 
planes (Fig.~\ref{fig1}) and the magnetic moment of each ion $(i)$ 
is represented by a classical vector, constrained by 
the polar angles $\theta_i$ and $\phi_i$. 
The spins are coupled via the Hamiltonian of Eqn.~\ref{eqn1}, 
for which periodic boundary conditions are applied. 
The zero-field spin configuration uses the results of 
elastic neutron scattering (see below) starting with collinear 
moments, antiferromagnetically coupled to all neighbors 
and oriented within the $xy-$planes. The simulation considered how the 
lowest-energy configuration of spins changes in a magnetic field. 
This was repeated for twenty-one evenly spaced angles $(\alpha)$ 
between the magnetic field and the hard-axis $(z)$ 
and for values in the range $0 \leq \alpha \leq 180^{\circ}$.
 
For a fixed value of $\alpha$, the magnitude of the field is 
increased from 0  to 60~T in $\mu_0\Delta H = 0.3$~T steps. 
Following the field increment, the energy of the cluster was minimized 
with respect to the angles $(\theta_n, \phi_m)$ using the Metropolis 
Monte Carlo algorithm~\cite{a46}. 
First, $\theta$ for one spin is changed to a new random value 
and the energy of the new configuration calculated from Eqn.~\ref{eqn1}.
If the energy decreases, the change is accepted. 
If the energy is raised by an amount $\Delta E$, the change is accepted 
with a probability $\exp(-\Delta E/k_{\rm B}T)$, otherwise 
the move is rejected. 
Next, the reorientation of the angle $\phi$ of the spin is considered 
in the same way as before applying the routine to the remaining seven 
moments in the cluster. This process is repeated 2000 times 
for a particular value of magnetic field. Once complete, the 
magnetization parallel to the applied field, 
${\bf M}(\alpha,{\bf H}).\frac{\bf H}{|{\bf H}|}$, was extracted and recorded. 
The powder average magnetization at a particular magnetic field, 
$M_{\rm avg}(H)$, is then determined by
\begin{equation}
M_{\rm avg}(H)=\frac{\sum_{\alpha}\left[ {\bf M}(\alpha,{\bf H}).\frac{\bf H}{|{\bf H}|}\right]
\sin \alpha \Delta \alpha}
{\sum_{\alpha} \sin \alpha \Delta \alpha}
\label{eqn2}
\end{equation}
Owing to the condition which allows the energy to be raised during a 
field step, the final spin configuration will be within an energy of 
$k_{\rm B}T$ from the true ground state. Hence, the parameter $T$ 
plays the role of temperature. Here, $T$ is chosen to be 0.1~K,
so as to find the most probable ground state spin configuration.

\subsection{DFT calculation of spin density}
The exchange-coupling constant $J$ can be related to the 
energy difference between states with different 
spin multiplicities~\cite{a47,b47,c47,d47}. 
For this purpose, accurate unrestricted wave functions for the 
ferromagnetic (FM) and AFM spin states are required. 
We have investigated the FM and several combinations of 
AFM states, imposing a coupling only along Ni---FHF---Ni 
or the Ni---pyz---Ni directions, or along both, in order to 
estimate $J$ and $J'$ as well as the total coupling.
 
The CRYSTAL14 code~\cite{a48} was used to perform DFT 
calculations with periodic boundary conditions on relevant 
FM and AFM phases of [Ni(HF$_2$)(pyz)$_2$]SbF$_6$ 
using the B3LYP hybrid functional. 
The basis set was $6-31G(d,p)$~[\onlinecite{a49}] 
for all of the atoms except for Sb which utilized 
$3-21G(d)$~[\onlinecite{a50}]. 
Electron-spin and charge-density maps were calculated using 
the routines in CRYSTAL14 and the spin-atomic distribution was 
determined using Mulliken partitioning~\cite{a51}. 

\begin{table*}[t]
\centering
\caption{ X-ray crystallographic details and refinement results 
for [Ni(HF$_2$)(pyz)$_2$]SbF$_6$. Here, $F$ = amplitude of structure 
factor of reflection and $\sigma$ = standard deviation of $F$. 
The $R$-factors are given by $R = \sum |F_{\rm obs}-F_{\rm calc}|/\sum |F_{\rm obs}|$, 
$R_{\rm int} = \sum |F_{\rm obs}^2-F_{\rm calc}^2|/\sum |F_{\rm obs}^2|$ 
and $wR = (\sum w|F_{\rm obs}-F_{\rm calc}|^2/\sum|wFobs|^2)^{\frac{1}{2}}$, 
where the sums run over all data points and $w$ is a weighting factor. $S$ 
is the goodness of fit. Errors in parentheses are one 
standard deviation. 
}
\label{table1}
\begin{tabular}{ll}
\hline\hline
Parameter (units)                                                &                                    \\
\hline
Facility/beamline                                                & Advanced Photon Source/15-ID-B     \\
Chemical formula                                                 & C$_8$H$_9$N$_4$F$_8$NiSb           \\
Crystal system                                                   & Tetragonal                         \\
Space group                                                      & $P4/nmm$ (\#129; origin choice 2)  \\
Temperature (K)                                                  & 15(2)                              \\
Wavelength, $\lambda$ (\AA)                                      & 0.38745                            \\
$a$, $b$, $c$ (\AA)                                              & 9.8797(2), 9.8797(2), 6.4292(1)    \\
Volume (\AA$^3$)                                                 & 627.54(2)                          \\
Formula units per unit cell, $Z$                                 & 2                                  \\
Absorption coefficient, $\mu$ (mm$^{-1}$)                        & 1.98                               \\
Crystal size ($\mu$m$^{3)}$                                          & $10 \times 10 \times 2$            \\
\hline    
{\bf Data collection}                                            &                                    \\
No. of measured reflections [$F > 3\sigma(F)$]                   & 74100                              \\
No. of independent reflections                                   & 2851                               \\
No. of observed reflections                                      & 2484                               \\
$R_{\rm int}$                                                    & 0.0780                             \\
\hline
{\bf Spherical-atom refinement}                                  &                                    \\
$R$ [$F > 3\sigma(F)$], $R_{\rm all}$, $wR$, $S$                 & 0.0318, 0.0427, 0.0612, 1.111      \\
No. parameters                                                   & 36                                 \\
\hline\hline
\end{tabular}
\end{table*}

\begin{figure}[t]
\centering
\includegraphics[width=8.8cm]{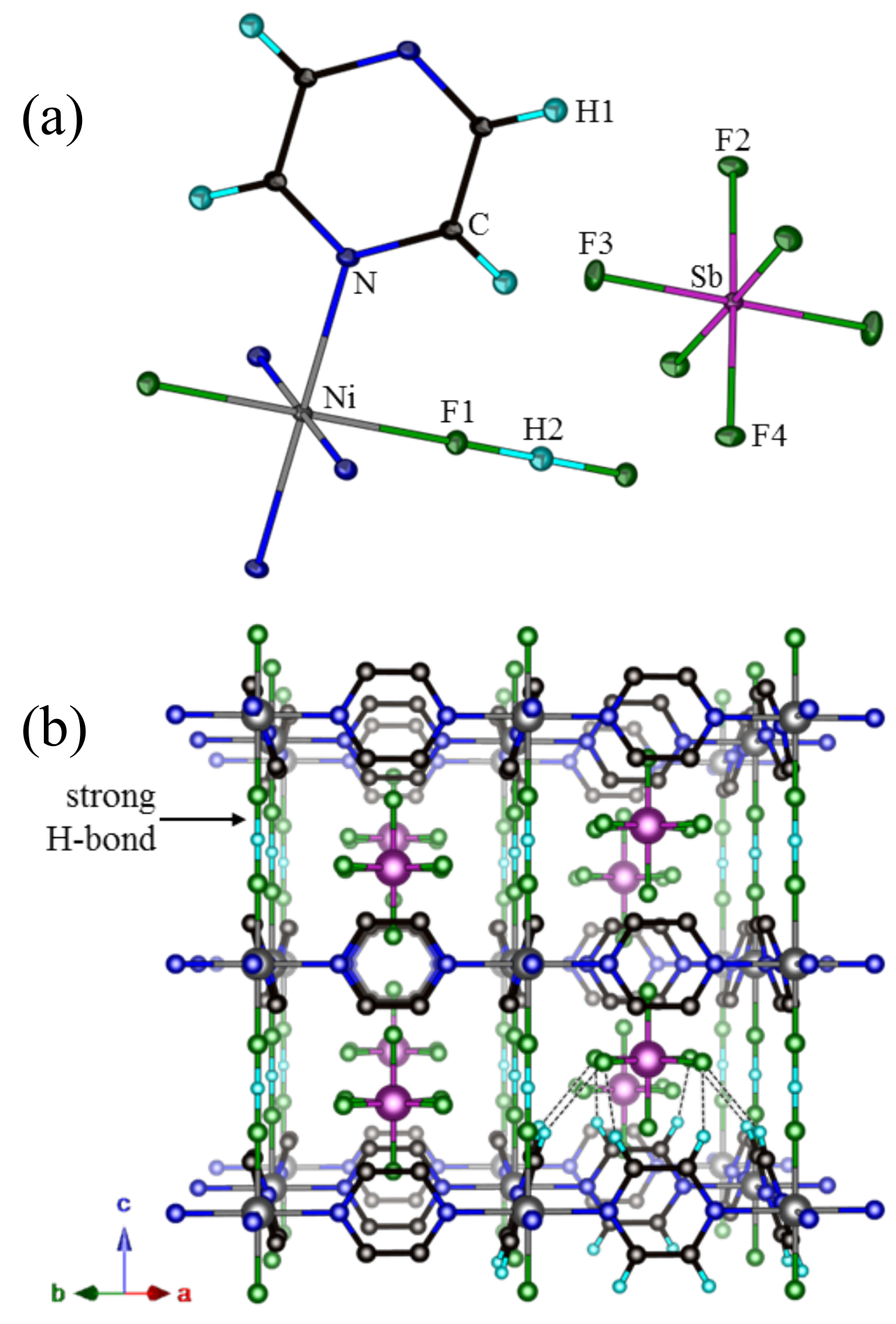}
\caption{Structure of [Ni(HF$_2$)(pyz)$_2$]SbF$_6$
determined at $T = 15$~K by microcrystal X-ray diffraction. 
(a)~Thermal ellipsoid plot (50~\% 
displacement parameters) showing the basic building blocks 
and atom-labeling scheme. Ni(II) ions have 4-fold rotational 
symmetry while the HF$_2^-$ constituent atoms H2 and F1 have 
respective site symmetries of $\bar{4}m2$ and $2mm$. 
(b)~Polymeric metal-organic framework including 
interstitial SbF$_6^-$ ions. Each Sb atom occupies the 
4-fold rotation axis that lies parallel to the $c$-direction but the centroid 
of the SbF$_6^-$ ion is displaced about 0.6~\AA
 relative to the ideal body-centered position. 
For clarity, only the lower right quadrant depicts the weak 
hydrogen bonds that exist between pyz ligands and the 
SbF$_6^-$ ion (H1$\cdots$F3 = 2.478~\AA; dashed lines). 
}
\label{fig2}
\end{figure}

\section{Results and discussion}
\subsection{Low-temperature chemical and magnetic structure}

\begin{table}[t]
\centering
\caption{Comparison of selected bond lengths (\AA) and bond angles (degrees) 
for [Ni(HF$_2$)(pyz)$_2$]SbF$_6$ as determined by the microcrystal 
X-ray and neutron powder diffraction studies.
(*Compared to $\Sigma_{\rm vdw}$ of 2.94~\AA ~for a pair of fluorine atoms.)} 
\label{table3}
\begin{tabular}{lrrr}
\hline\hline
                    & X-rays              & Neutrons             & Neutrons \\
                    & ($T = 15$ K)        & ($T = 20$ K)         & ($T = 1.5$ K) \\
\hline
Ni-N                & 2.098(1)            & 2.095(2)             & 2.096(2)  \\
Ni-F1               & 2.076(1)            & 2.067(6)             & 2.070(6)  \\
H2$\cdots$F1        & 1.138(1)            & 1.149(6)             & 1.146(6)  \\
F1$\cdots$F1*       & 2.276(1)            & 2.297(9)             & 2.292(9)  \\
C-N                 & 1.338(1)            & 1.338(3)             & 1.338(3)  \\
C-C                 & 1.391(2)            & 1.403(3)             & 1.402(3)  \\
C-H1/D1             & 0.95                & 1.071(4)             & 1.071(4)  \\
Sb-F2               & 1.869(2)            & 1.889(4)             & 1.890(4)  \\
Sb-F3               & 1.886(1)            & 1.892(11)            & 1.893(11) \\
Sb-F4               & 1.863(2)            & 1.771(12)            & 1.770(13) \\
Ni$\cdots$Ni ($c$-axis)    & 6.4292(1)           & 6.4319(2)            & 6.4318(2) \\
Ni$\cdots$Ni [1 1 0; 1 $\bar{1}$ 0] & 6.9860(2)           & 6.9958(1)            & 6.9956(1) \\
                    &                     &                      &           \\
F1-Ni-F1            & 180                 & 180                  & 180       \\
N-Ni-N              & 90, 180             & 90, 180              & 90, 180   \\
F1-Ni-N             & 90                  & 90                   & 90        \\
Ni-F1$\cdots$H2          & 180                 & 180                  & 180       \\
Ni-N$\cdots$N            & 180                 & 180                  & 180       \\
C-N-C               & 116.97(9)           & 116.8(3)             & 116.8(3)  \\
N-C-H1/D1           & 119.2               & 119.1(4)             & 119.0(4)  \\
F2-Sb-F3            & 89.27(3)            & 89.8(4)              & 89.8(5)   \\
F2-Sb-F4            & 180                 & 180                  & 180       \\
F1-Ni-N-C           & 73.04(4)            & 73.7(1)              & 73.7(1)   \\
\hline\hline
\end{tabular}
\end{table}

\noindent
{\it Microcrystal X-ray diffraction}.
The 15~K structure (Fig.~\ref{fig2}) of [Ni(HF$_2$)(pyz)$_2$]SbF$_6$ 
was solved in the tetragonal space group $P4/nmm$ based on the 
single-crystal X-ray diffraction data. 
Full details of the structural refinement are given in Table~\ref{table1}.
Selected bond lengths and bond angles can be found 
in Table~\ref{table3}. Each Ni(II) ion is axially-coordinated to 
two F1 atoms at a distance of 2.076(1)~\AA
~[the atom labelling scheme is shown in Fig.~\ref{fig2}(a)]. 
The F1 atoms belong to bridging HF$_2^-$ ligands that 
form one-dimensional linear Ni---FHF---Ni chains along the $c$-axis 
with respective F1$\cdots$F1 and Ni$\cdots$Ni separations of 2.276(1)~\AA
~and 6.4292(1)~\AA. These linkages mediate an intrachain interaction 
through $\sigma$-bond magnetic coupling as established by 
experiment and DFT (see below). Pyrazine ligands join the Ni(II) ions 
[Ni---N = 2.098(1)~\AA] along the $[1 1 0]$ and $[1 \bar{1} 0]$ 
directions to produce two-dimensional (2D) square sheets in 
the $ab$-plane [Fig.~\ref{fig2}(b)], with equal Ni---Ni separations 
of 6.9860(2)~\AA, and provide the interchain interactions. 
The slight difference in Ni---F1 and Ni---N bond lengths results in 
a weakly compressed octahedral NiN$_4$F$_2$ coordination 
environment. Trans-coordinated pyz ligands are counter-rotated 
and tilt away from the NiN$_4$ plane by $73.04(4)^{\circ}$. 
The ordering of the pyz ligands in [Ni(HF$_2$)(pyz)$_2$]SbF$_6$ 
contrasts the two-fold positional disorder encountered in 
the related quasi-two-dimensional (Q2D) layered 
coordination polymers Ni$Z$$_2$(pyz)$_2$ ($Z$ = Cl, Br, I, NCO), 
which crystallize in the $I4/mmm$ space group~\cite{a52,a53}, 
The structurally related material, Ni(NCS)$_2$(pyz)$_2$, has 
monoclinic symmetry and no apparent pyz disorder~\cite{a52}.
 
\begin{table*}[t]
\centering
\caption{Refinement details from neutron powder-diffraction 
data for [Ni(HF$_2$)(pyz-d$_4$)$_2$]SbF$_6$ at 1.5~K 
and 20~K. The magnetic propagation vector in reciprocal lattice units (r.l.u.) is $k$ 
and $\mu$ is the refined Ni(II) magnetic moment. 
The goodness-of-fit parameters 
$R_{\rm F} = \sum|I_{\rm obs}^{\frac{1}{2}}-I_{\rm calc}^{\frac{1}{2}}|/\sum I_{\rm obs}^{\frac{1}{2}}$, 
$R_{\rm Bragg} = \sum |I_{\rm obs}-I_{\rm calc}|/\sum |I_{\rm obs}|$, 
where $I_{\rm obs}$ is the observed intensity, $I_{\rm calc}$ 
is the calculated intensity and the sum runs over all data points. 
$R_{\rm mag}$ is the equivalent $R$ factor to $R_{\rm Bragg}$ 
applied to the fit of the magnetic scattering only. 
}
\label{table2}
\begin{tabular}{lll}
\hline\hline
Parameter (units)   &                                 &                                              \\
\hline
Formula             & C$_8$HD$_8$N$_4$F$_8$NiSb       & C$_8$HD$_8$N$_4$F$_8$NiSb                   \\
Temperature (K)     & 20                              & 1.5                                         \\
Space group         & $P4/nmm$                        & $P4/nmm$                                    \\
$a$, $b$, $c$ (\AA) & 9.8936(1), 9.8936(1), 6.4319(2) & 9.8933(1), 9.8933(1), 6.4318(2)             \\
Volume (\AA$^3$)    & 629.57(2)                       & 629.53(2)                                   \\
$Z$                 & 2                               & 2                                           \\
$k$ (r.l.u.)        & --                              & (0, 0, 1/2)                                 \\
Order type          & --                              & G-type                                      \\
$\mu (\mu_{\rm B})$              & --                              & 2.03(7)                                     \\
WISH detector banks & Bank 2 + Bank 9                 & Bank 2 + Bank 9                             \\
$R_{\rm F}$ (\%)    & 5.42                            & 5.46                                        \\
$R_{\rm Bragg}$ (\%) & 3.64                           & 3.62                                        \\
$R_{\rm mag}$ (\%)   & --                             & 5.89 ($m\perp c$), 18.6 ($m \parallel c$)   \\
\hline\hline
\end{tabular}
\end{table*}

\begin{figure*}[t]
\centering
\includegraphics[width=17cm]{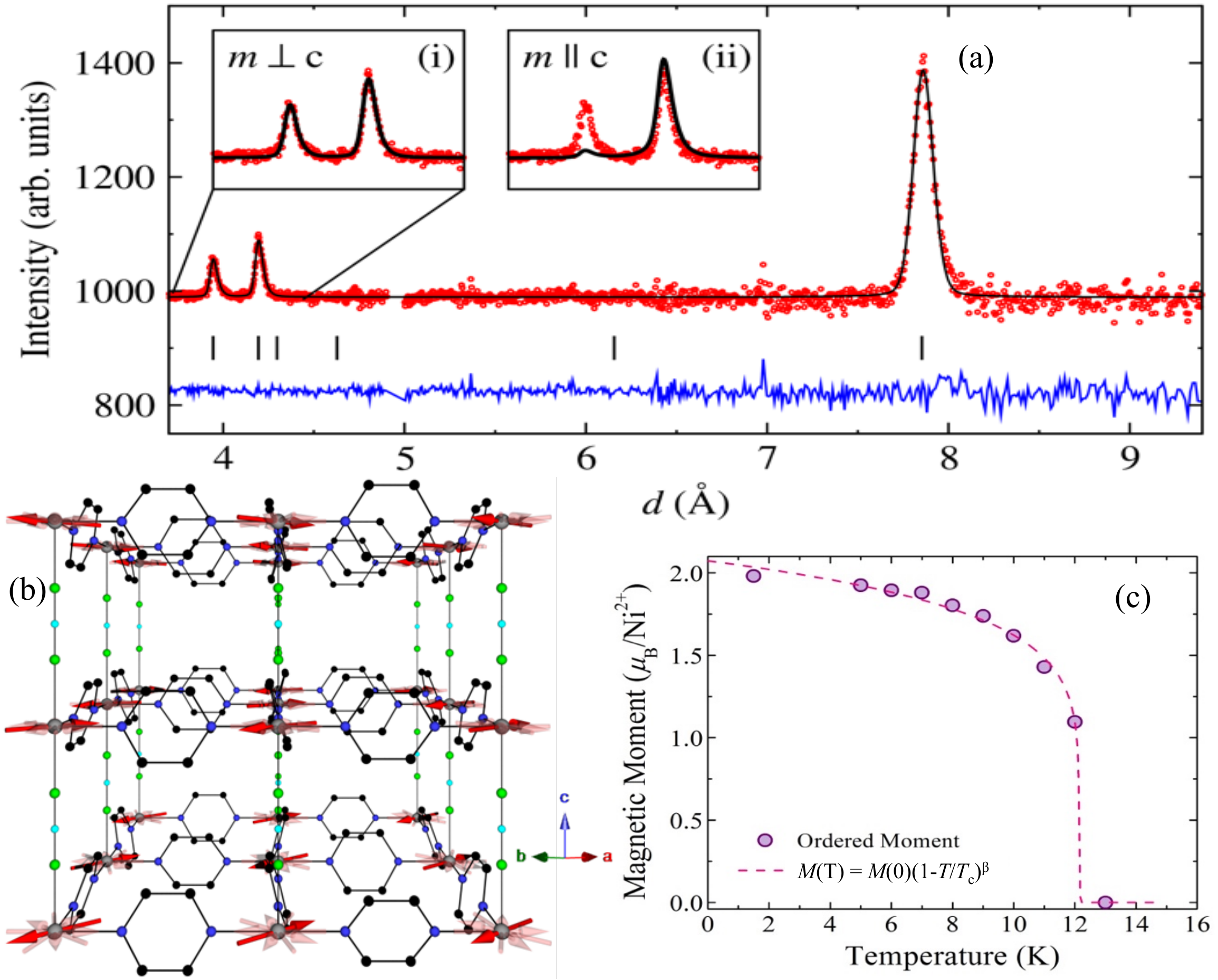}
\sloppypar
\caption{ Elastic neutron scattering data. (a)~Magnetic diffraction pattern (red points) 
for [Ni(HF$_2$)(pyz-d$_4$)$_2$]SbF$_6$ obtained by subtracting data 
collected at 20~K from that collected at 1.5~K (see Supplementary
Information (SI)~\cite{SI}). 
The fitted spectrum (black line) has the Ni(II) moment lying in the $ab$-plane. 
Bragg peaks are indicated by ticks and the blue line is the difference between 
the data and the fit. The insets show a comparison of the model calculated 
with the moments perpendicular~(i) or parallel~(ii) to the $c$-axis. 
(b) Zero-field magnetic structure (omitting pyz Hs and SbF$_6^-$). 
Collinear $XY-$ordered Ni(II) magnetic moment vectors are indicated 
by red arrows. (c)~$T-$dependence of the ordered Ni(II) magnetic moment. 
The power law fit yields~\cite{nonneel} $T_{\rm c} = 12.13(7)$~K and $\beta = 0.141(1)$. 
}
\label{fig3}
\end{figure*}

Interstitial sites within the [Ni(HF$_2$)(pyz)$_2$]$^+$ 
framework are occupied by charge-compensating SbF$_6^-$ ions. 
Significant close contacts of 2.478~\AA  
~exist between pyrazine H-atoms 
and equatorial Fs from the SbF$_6^-$ [Fig.~\ref{fig2}(b); dashed lines]. 
These weak C---H$\cdots$F hydrogen bonds probably constrain the pyz 
ligands to a single configuration and are unlikely to contribute 
to any additional magnetic exchange mechanism. 

\vspace{3mm}
\noindent
{\it Elastic neutron scattering.}
The chemical structure of [Ni(HF$_2$)(pyz-d$_4$)$_2$]SbF$_6$ 
was refined at 1.5 and 20~K using Rietveld analysis as implemented in 
FULLPROF (Fig. S1)~[\onlinecite{a42}]. 
Structural parameters derived from the microcrystal X-ray study 
of the hydrogenated phase were used as initial input. 
As anticipated, the deuterated material was found to be isostructural to the 
hydrogenated phase and the magnetic properties of the two compounds 
were found to be very similar based on magnetic susceptibility measurements. 
Unit-cell parameters derived from neutron scattering
can be found in Table~\ref{table2}, whereas
Table~\ref{table3} compares bond lengths and bond angles
provided by the X-ray and neutron experiments. 

Examining the difference in scattered neutron intensity 
obtained at 1.5 and 20~K [Fig.~\ref{fig3}(a)] reveals three distinct 
Bragg peaks at approximately 3.94~\AA, 4.19~\AA, and 7.85~\AA 
~that do not overlap any nuclear peaks. 
These are attributed to long-range antiferromagnetic order of 
Ni(II) moments in the material. Indexing of the superlattice peak at 7.85~\AA 
~requires doubling of the chemical unit cell along the $c$-axis. 
This indicates that the intrachain interaction along the HF$_2^-$ 
bridge is AFM in nature, as was also found to be the case in the 
isostructural Cu(II) and Co(II) congeners~\cite{a18}. 
Thus, the magnetic unit cell corresponds to a propagation 
vector ${\bf k} = (0, 0, \frac{1}{2})$ referenced to reciprocal lattice vectors 
based on the chemical unit cell.
 
Symmetry analysis and structure-factor calculations (see Fig. S1) 
demonstrated that the $d$-spacing positions of the magnetic-diffraction 
intensities were only consistent with a magnetic structure comprised 
of collinear spins anti-parallel to their nearest-neighbors (G-type). 
Two spin directions are then unique by symmetry:

\noindent
(i)~spins orthogonal to the crystallographic $c$-axis; or 

\noindent
(ii)~spins parallel to the $c$-axis. 

\noindent
These two scenarios were tested through refinement of the respective 
magnetic structure models against the diffraction data. 
It was found that only scenario (i) quantitatively predicts 
the relative intensities of all of the measured magnetic 
Bragg peaks, as is clearly shown in Fig.~\ref{fig3}(a), 
and corroborated by the respective agreement factors $(R_{\rm mag})$
of 5.89~\% and 18.6~\%. 
General comparison between calculated $(F_c^2)$ 
and observed $(F_o^2)$ structure factors for preferred 
configuration (i) show very good correlation (Table~\ref{table4}). 
We therefore conclude that 
[Ni(HF$_2$)(pyz)$_2$]SbF$_6$ exhibits the 3D $XY$-AFM ground state 
depicted in Fig.~\ref{fig3}(b).
 
At $T=1.5$~K, the experimentally determined Ni(II) magnetic moment 
has a magnitude of $2.03(7)\mu_{\rm B}$, which is very close to the full 
moment expected for an $S = 1$ ion, $gS\mu_{\rm B} = 2.08\mu_{\rm B}$, 
given the published~\cite{a18} powder-average $g$-factor. 
This observed full moment precludes strong quantum 
fluctuations in the ground state which were prominent in the 
Q2D Heisenberg $S = \frac{1}{2}$ Cu(II) congener, for which a 
reduced ordered moment of $0.6(1)\mu_{\rm B}$ was found~\cite{a20}. 
The differing results can be attributed to the smaller 
spin-quantum number and strong quantum fluctuations, 
significantly larger spatial exchange anisotropy, 
and the lack of single-ion anisotropy in the copper material.

\renewcommand{\arraystretch}{1.3}
\begin{table}[]
\centering
\caption{Observed and calculated nuclear and magnetic structure factors 
for [Ni(HF$_2$)(pyz)$_2$]SbF$_6$ as determined by the 1.5~K 
neutron-diffraction study. $F_{\rm o}$ and $F_{\rm c}$ 
correspond to observed and calculated structure values, 
respectively, for a given $d$-spacing.}
\label{table4}
\begin{tabular}{cccccccc}
\hline\hline
$h$  & $k$ & $l$ &  & $F_{\rm o}^2$ & $F_{\rm c}^2$ ($\perp c$) & $F_{\rm c}^2$ ($\parallel c$) & $d$-spacing (\AA) \\
\hline
1 & 0 & $\frac{3}{2}$ &  & 2.999 & 2.927                    & 0.431     & 3.9358        \\
1 & 2 & $\frac{1}{2}$ &  & 3.539 & 3.662                    & 5.079     & 4.1856        \\
1 & 0 & $\frac{1}{2}$ &  & 2.631 & 2.854                    & 2.241     & 7.8454        \\
\hline\hline
\end{tabular}
\end{table}

\begin{figure}[t]
\centering
\includegraphics[width=8.5cm]{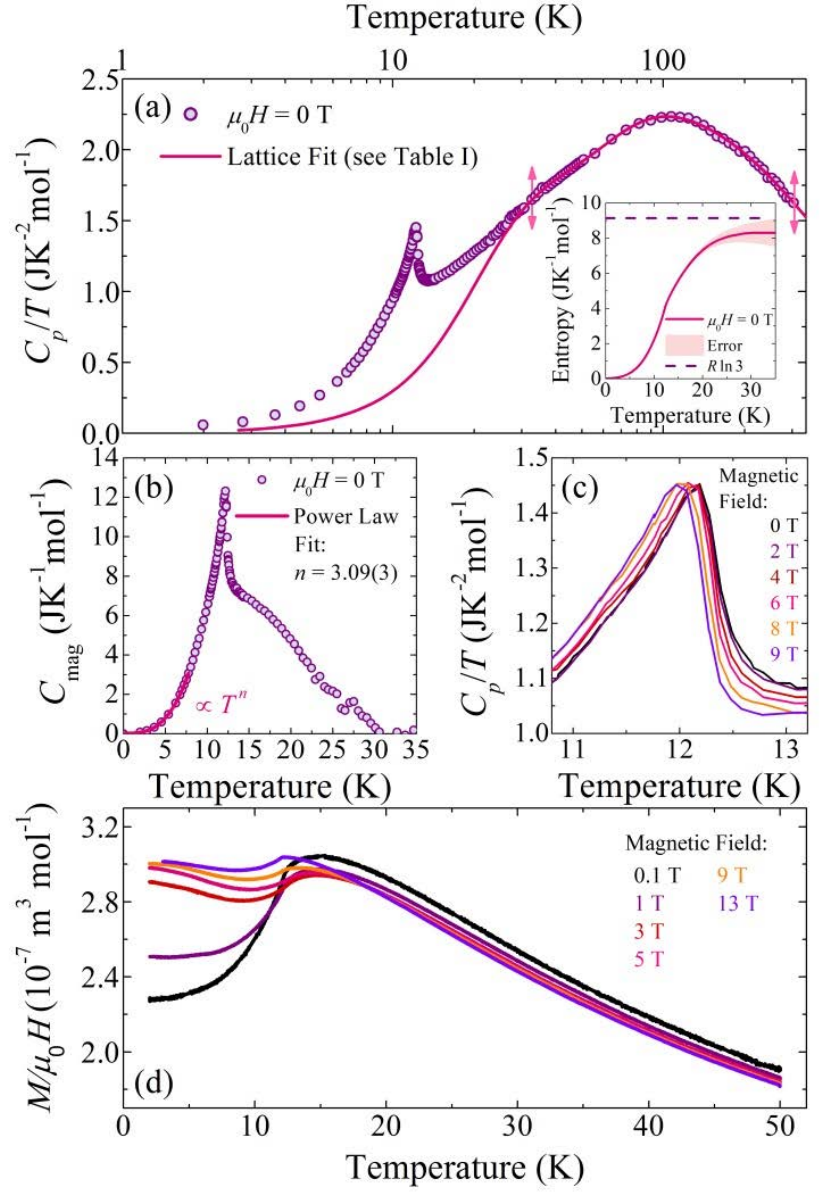}
\sloppypar
\caption{(a)~Ratio of heat capacity to temperature $T$ for [Ni(HF$_2$)(pyz)$_2$]SbF$_6$ 
(points). Data for $T > 32$~K have been fitted (arrows) to a model of one 
Debye mode and two Einstein modes (line). The inset shows the magnetic entropy 
up to 35~K where it can be seen to approach the expected value of $R\ln 3$ 
for $S = 1$ ions. (b)~Magnetic heat capacity $(C_{\rm mag})$ exhibits a 
broad maximum preceding a sharp transition at 12.2(1)~K, 
indicating the onset of long-range order. Data below 8~K were fitted to a 
power law, $T^n$, yielding $n = 3.09(3)$. 
(c)~Field dependence of $T_{\rm c}$ for magnetic fields $0 \leq \mu_0 H \leq 9$~T. 
(d)~Field-dependent linear susceptibility $(M/\mu_0 H)$ versus 
$T$ for [Ni(HF$_2$)(pyz)$_2$]SbF$_6$, showing a broad maximum, 
the temperature of which is suppressed by increasing field.
}
\label{fig4}
\end{figure}

Data plotted in Fig.~\ref{fig3}(c) were obtained by fitting the ordered 
moment to each measured diffraction pattern. 
The resulting fit of these data to a power law of the form~\cite{nonneel}, 
$M(T) = M(0)\left[1 - T/T_{\rm c}\right]^{\beta}$,  yielded
$T_{\rm c} = 12.13(7)$~K and $\beta= 0.141(1)$. 
For most systems the critical region in which power-law behavior 
applies is restricted to $1-T/T_{\rm c} < 10^{-2}$~[\onlinecite{a54}]. 
Sparse data in the vicinity of $T_{\rm c}$ suggests caution 
in assigning $\beta$ to a particular model. 

\vspace{3mm}
\noindent
{\it Field-dependent heat capacity and susceptibility.}
The zero-field heat capacity plotted as $C_p/T$ versus $T$ 
[Fig.~\ref{fig4}(a)] shows a sharp maximum at 12.2(1)~K, 
indicating a transition to long-range order in addition to a sloping 
background resulting from the phonon contribution. 
The data for $T \geq 32$~K have been modeled~\cite{a24} with one 
Debye and two Einstein phonon modes and the resulting fit parameters 
are tabulated in Table~\ref{table5}. 
The three lattice modes show similar energy scales to 
those deduced from analogous analysis of the copper~\cite{a24} 
and cobalt~\cite{a20} isomorphs, which results from the 
shared structure of this family of compounds. 
Subtracting the phonon heat capacity from the total measured 
heat capacity of [Ni(HF$_2$)(pyz)$_2$]SbF$_6$, 
the magnetic heat capacity shows a broad hump that develops on 
cooling and precedes a transition to long-range order (Fig.~\ref{fig4}). 
This broad feature corresponds to a significant 
reduction in the spin entropy for $T > T_{\rm c}$ (inset) and is 
likely to result from the combination of two mechanisms that 
restrict the magnetic degrees of freedom of the system: 
(i)~development of $XY$-anisotropy of the individual Ni(II) moments, 
as well as (ii)~the build-up of AFM spin correlations among 
neighboring Ni(II) ions dispersed along the Ni---FHF---Ni chains.

\begin{table}[]
\centering
\caption{Fitted parameters for the phonon heat capacity of 
[$M$(HF$_2$)(pyz)$_2$SbF$_6$, where $M$ = Ni(II), Cu(II)~[\onlinecite{a24}] 
and Co(II)~[\onlinecite{a20}]. The simplest lattice model~\cite{a24} 
required to fit the data included one Debye (D) and two Einstein (E) 
modes that are each determined by an amplitude $(A_i)$ and characteristic 
temperature $(\theta_i)~(i = $D, E).}
\label{table5}
\begin{tabular}{lllll}
\hline\hline
                                     & Ni(II)   & Cu(II) & Co(II) &  \\
\hline
$A_{\rm D}$ (JK$^{-1}$mol$^{-1}$)    & 123(3)   & 76(1)  & 82(1)  &  \\
$\theta_{\rm D}$ (K)                 & 148(3)   & 94(8)  & 80(1)  &  \\
$A_{\rm E1}$ (JK$^{-1}$mol$^{-1}$)   & 271(6)   & 134(4) & 174(3) &  \\
$\theta_{\rm E1}$ (K)                & 345(7)   & 208(4) & 177(2) &  \\
$A_{\rm E2}$ (JK$^{-1}$mol$^{-1}$)   & 240(6)   & 191(4) & 287(3) &  \\
$\theta_{\rm E2}$ (K)                & 860(30)  & 500(3) & 448(6) &   \\
\hline\hline
\end{tabular}
\end{table}

For $T\ll T_{\rm c}$, spin-wave excitations are the dominant 
contribution to the magnetic heat capacity. 
The data below 8~K have been represented with a power law $T^n$, 
where $n$ has a fitted value of $n = 3.09(3)$. 
This exponent was previously reported for the total sample 
heat capacity~\cite{a18}, but the new results specifically 
exclude contributions to the measurement from the 
Debye mode. A $T^3$ dependence of the heat capacity is expected 
for an AFM ordered system within which the magnetic 
excitations (magnons) propagate in three dimensions~\cite{a55}, 
which highlights the need to include the effects of $J$ and $J'$ 
in the analysis of the spin-wave excitation spectra measured 
from inelastic neutron data (see below).

\begin{figure}[t]
\centering
\includegraphics[width=8.5cm]{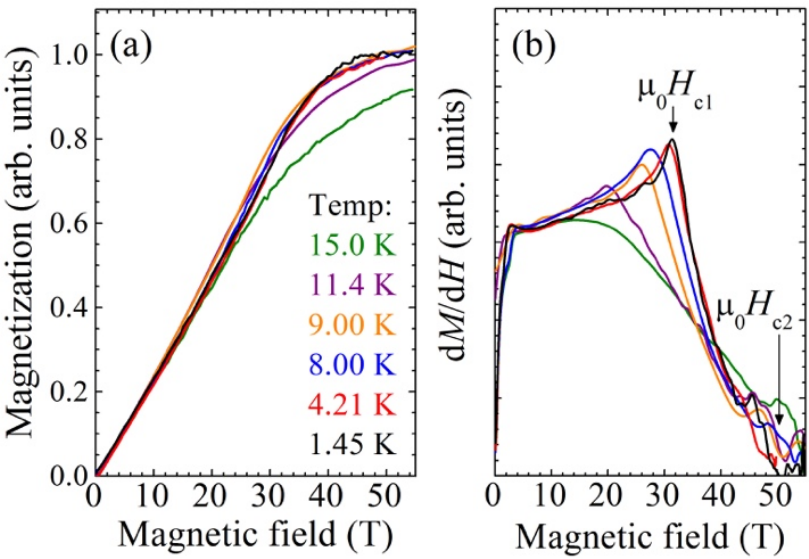}
\caption{(a)~Magnetization $M$ versus applied magnetic field
$\mu_0H$ for [Ni(HF$_2$)(pyz)$_2$]SbF$_6$ measured in pulsed fields 
for selected $T \leq 15$~K. (b)~Differential susceptibility
${\rm d}M/{\rm d} H$ versus $\mu_0 H$; such data were 
used to determine the $T$ dependence of the critical 
field $H_{\rm c1}$. The values of $H_{\rm c1}$ and $H_{\rm c2}$ 
at 1.45~K are shown by arrows.}
\label{fig5}
\end{figure}

The field dependence of the heat capacity [$\mu_0 H \leq 9$~T; 
Fig.~\ref{fig4}(c)] and the linear susceptibility 
[$M/\mu_0 H;~ \mu_0H \leq 13$~T; Fig.~\ref{fig4}(d)] 
implies that the transition temperature is suppressed in applied magnetic field, 
as required for an AFM ground state. 
The overall field dependence is weak for fields of up to 13~T, 
owing to the large critical fields in this system, and the 
high-field portion of the phase diagram was explored using 
pulsed-field magnetization measurements (see below). 

\subsection{Experimental determination of  
$D$, $J$ and $J'$}

\noindent
{\it High-field magnetization.}
The pulsed-field magnetization of [Ni(HF$_2$)(pyz)$_2$]SbF$_6$ 
at low temperatures shows a slightly concave rise with increasing 
field and a broadened approach to saturation [Fig.~\ref{fig5}(a)]. 
Two critical fields are identified at 1.45~K, which correspond to an 
initial increase and subsequent decrease in ${\rm d}M/{\rm d}H$ 
close to 30~T and the point where ${\rm d}M/{\rm d}H\rightarrow 0$ 
near 50~T [Fig.~\ref{fig5}(b), arrows]. 
To interpret the primary features, we adopt a mean-field 
model for AFM-coupled easy-plane $S = 1$ ions which may be justified 
based on the ground state determined from elastic neutron 
scattering (see above). Within this model the saturation field 
should be anisotropic with an easy plane and a hard axis such that 
two saturation fields could be observed in the powder data. 
For fields perpendicular and parallel to the magnetic hard axis, 
each of these saturation fields ($H_{\rm c1}$ and $H_{\rm c2}$, 
respectively) can be calculated by
\begin{equation}
\mu_0H_{\rm c1} = \frac{2n\langle J\rangle}{g\mu_{\rm B}}
\label{eqn3}
\end{equation}
and
\begin{equation}
\mu_0H_{\rm c2} = \frac{2n\langle J\rangle+D}{g\mu_{\rm B}}
\label{eqn4}
\end{equation}
where $n$ is the number of nearest neighbors for each magnetic 
ion and $\langle J\rangle$ is an average spin-exchange interaction. 
In a powder measurement of the differential susceptibility, 
a decrease in ${\rm d}M/{\rm d}H$ occurs at $H_{\rm c1}$ 
once a portion of the sample is saturated. Then,
${\rm d}M/{\rm d}H$ continues to decrease to $H_{\rm c2}$, 
where no further increase in the magnetization can occur as
all Ni(II) moments are aligned with the magnetic field.
 
\begin{figure}[t]
\centering
\includegraphics[width=8.5cm]{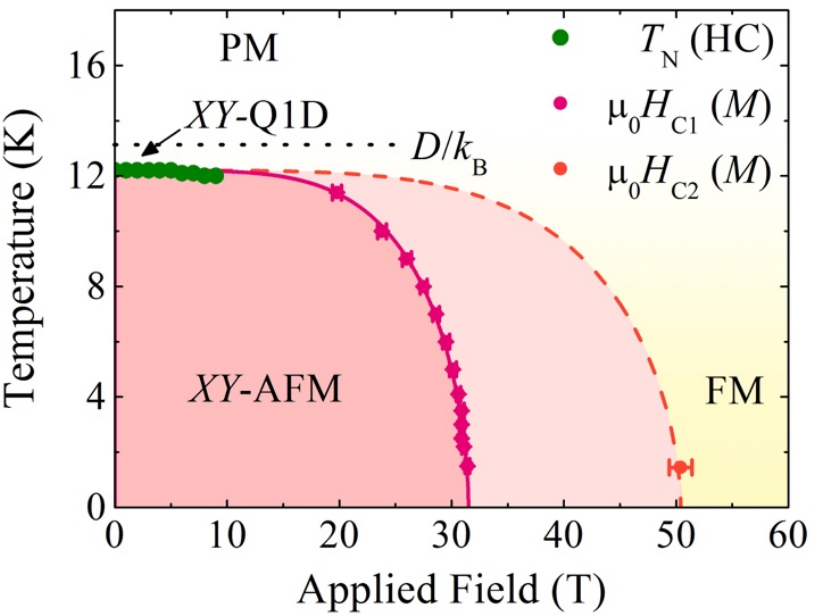}
\caption{Field-temperature $(H,T)$ phase diagram for 
[Ni(HF$_2$)(pyz)$_2$]SbF$_6$, mapped out using data from 
heat capacity (HC) and pulsed-field magnetization (M) measurements. 
Here, $XY$-AFM = long-range $XY$-antiferromagnetic order, 
FM = fully polarized phase; and $XY$-Q1D is a region where the moments 
are antiferromagnetically correlated along the chains, with the 
Ni(II) moments oriented perpendicular to those chains; $H_{\rm c1}~(H_{\rm c2})$ 
is the field at which the magnetization saturates when the magnetic 
field is applied perpendicular (parallel) to the hard axis. 
The $\mu_0 H_{\rm c1}$ phase boundary (solid line) is modeled using 
Eqn~\ref{eqn5}, whilst the $\mu_0 H_{\rm c2}$ boundary (dashed line) 
is a guide to the eye.  The energy scale of the single-ion anisotropy $(D)$ 
is indicated by a dotted line for reference.
}
\label{fig6}
\end{figure}

The first critical field, $H_{\rm c1}$ for [Ni(HF$_2$)(pyz)$_2$]SbF$_6$ 
is easily identified at all $T < 15$~K (Fig.~\ref{fig5}). 
Combining these critical fields with the heat-capacity results (above), 
the field-temperature phase diagram can be derived (Fig.~\ref{fig6}). 
The phase boundary for fields applied perpendicular to the 
magnetic hard-axis (solid lines) is fitted to the expression
\begin{equation}
T_{\rm c}(H)=T_{\rm c} (0) 
\left[1-\left(\frac{H}{H_{\rm c1}}\right)^{\alpha_1}\right]^{\beta_1}.
\label{eqn5}
\end{equation}
Fixing $T_{\rm c}(0) =12.2$~K, the resultant fitted parameters are
$\mu_0 H_{\rm c1} = 31.5(2)$~T, $\alpha_1 = 4.6(4)$,
$\beta_1 = 0.56(4)$.
 
The temperature evolution of the saturation field for 
fields parallel to the magnetic hard axis $(\mu_0H_{\rm c2})$ 
is harder to follow; the ${\rm d}M/{\rm d}H$ signal is lowest in 
this field region and arises from a diminishing proportion 
of the sample as this field is approached. 
In addition, as the temperature increases, the transition to 
saturation is broadened further. At the lowest temperatures, 
however, the saturation field can be identified and is 
found to be $\mu_0 H_{\rm c2} = 50.4(2)$~T at 1.45~K.
 
Using these critical fields in conjunction with Eqns.~\ref{eqn3}
and \ref{eqn4} and the powder average $g$-value~\cite{a18} of 2.08,
we determine $D = 13.2(5)$~K and 
$n\langle J\rangle\equiv 2J + 4J' = 22.0(2)$~K. 
To decompose $n\langle J\rangle$ into individual $J$ and $J'$ contributions, 
we appeal first to a catalog of related coordination polymers that 
also contain square [Ni(pyz)$_2$]$^{2+}$ motifs~\cite{a52,a56}. 
From Table~\ref{table6}, we glean an average $J' = 0.9(2)$~K. 
Applying this to [Ni(HF$_2$)(pyz)$_2$]SbF$_6$ leads to $J = 9.2(5)$~K. 
Therefore, the resultant $D/J$ and $J'/J$ ratios are more than and 
less than one, respectively, which is consistent with a preliminary 
DFT study~\cite{a18} used to calculate $J'$. 
These parameters are in good agreement with the results of the 
INS measurements and the Monte-Carlo simulations detailed below.

\begin{table}[]
\centering
\caption{Review of the exchange interaction strengths 
mediated through Ni-pyz-Ni linkages $(J')$ for coordination polymers 
consisting of approximately square [Ni(pyz)$_2$]$^{2+}$ 
plaquettes similar to those found in [Ni(HF$_2$)(pyz)$_2$]SbF$_6$. 
The values listed were obtained by thermodynamic measurements. }
\label{table6}
\begin{tabular}{llll}
\hline\hline
System                     & $J'$ (K)       & $T_{\rm c}$ (K) & Ref. \\
\hline
NiCl$_2$(pyz)$_2$          & 0.49(1)        & $<0.08$        & \onlinecite{a52}   \\
NiBr$_2$(pyz)$_2$          & 1.00(5)        & 1.8(1)         & \onlinecite{a52}   \\
NiI$_2$(pyz)$_2$           & $<1.19$        & 2.5(1)         & \onlinecite{a52}   \\
Ni(NCS)$_2$(pyz)$_2$       & 0.82(5)        & 1.8(1)         & \onlinecite{a52}   \\
$[$Ni(H$_2$O)$_2$(pyz)$_2]$(BF$_4$)$_2$ & 1.05(5)        & 3.0(1)         & \onlinecite{a56}   \\
\hline\hline
\end{tabular}
\end{table}

\begin{figure}[t]
\centering
\includegraphics[width=8.5cm]{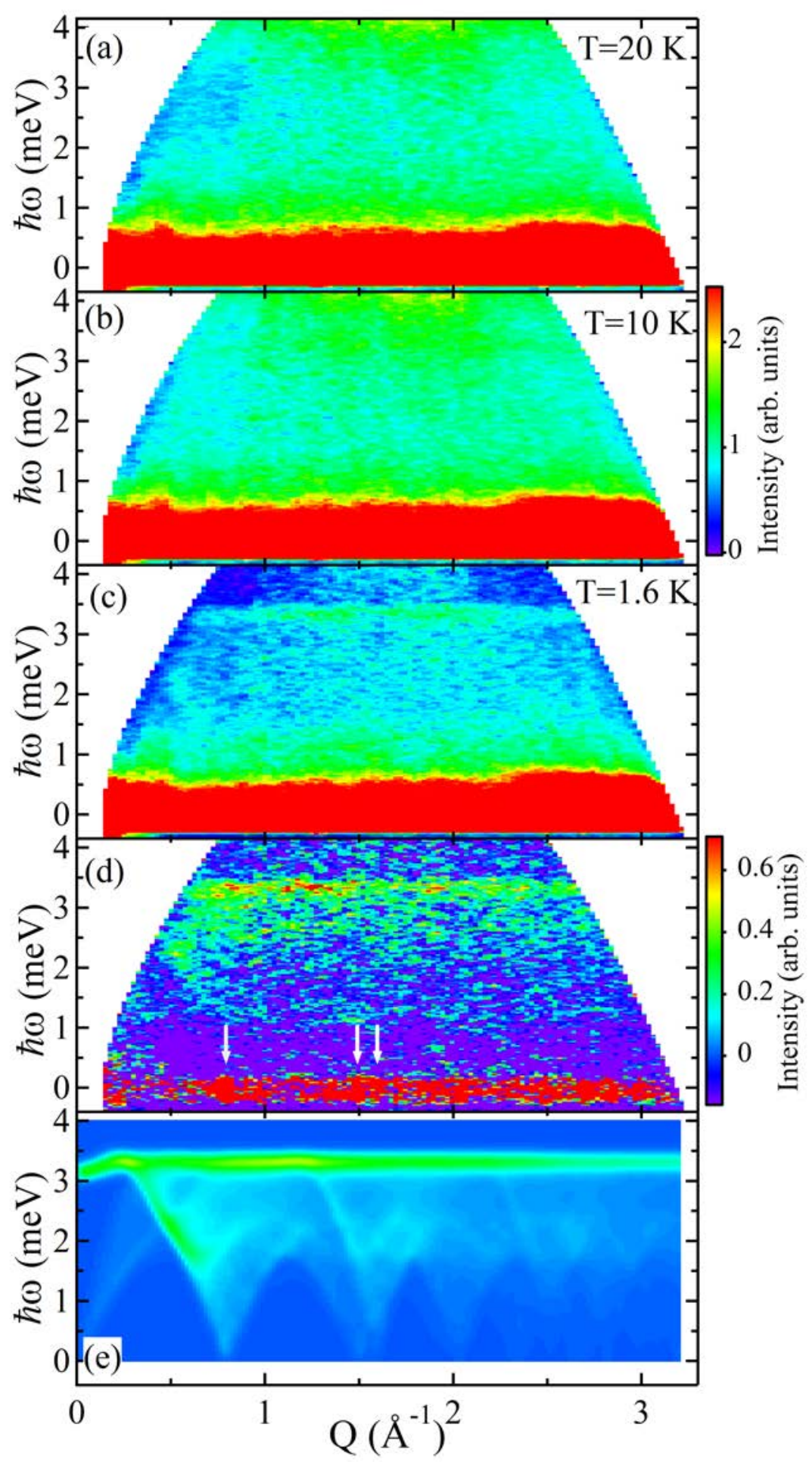}
\caption{Intensity contour plots of INS data with no background subtraction for:  
(a)~$T = 20$~K, (b)~$T = 10$~K, and (c)~$T = 1.6$~K. 
Panel (d) is the $T = 1.6$~K measurement after subtracting the $T = 20$~K data as a 
background. The data were binned in units of 0.025~meV and 0.025~\AA$^{-1}$
and then smoothed once with a $3 \times 3$-bin-sized Gaussian kernel. 
White arrows in panel (d) highlight the locations of prominent magnetic 
Bragg peaks as validated by the diffraction experiment on WISH 
[see Fig.~\ref{fig3}(a)]. Panel (e) shows the simulated powder INS spectrum 
for the parameters  $D = 13.3(1)$~K, $J = 10.4(3)$~K and $J' = 1.4(2)$~K. 
}
\label{fig7}
\end{figure}

Using these results, the form of the phase diagram can be interpreted. 
On cooling the sample from room temperature in zero-field, the sample 
moves from a paramagnetic (PM) phase to a region within which the 
Ni(II) moments develop $XY$-anisotropy ($XY$-Q1D) and their 
directions are antiferromagnetically correlated for neighboring ions 
along the Ni---FHF---Ni chains, with the moments arranged perpendicular 
to these chains. On cooling further, there is a magnetic phase transition 
to an ordered state. We assign this to long-range order with $XY$ moments 
antiferromagnetically and collinearly aligned to their nearest neighbors. 
Starting from this ordered phase and applying a magnetic field, 
the system is driven through a field-induced phase transition to a FM-like phase. 
This occurs for fields bounded by the range 
$H_{\rm c1} \leq H \leq H_{\rm c2}$, depending on whether the field is 
applied perpendicular $(H_{\rm c1})$ or parallel $(H_{\rm c2})$ to the 
$z$-direction. For powder measurements of the magnetization this 
anisotropy leads to the slow and broad approach toward saturation.

\vspace{3mm}
\noindent
{\it Inelastic neutron scattering.} 
Figure~\ref{fig7}(a-c) shows the measured powder INS energy-momentum 
transfer spectrum of [Ni(HF$_2$)(pyz-d$_4$)$_2$]SbF$_6$ 
in zero-field for $T = 1.6, 10$ and 20~K. As the temperature decreases an upper 
bound of the spectrum appears at a neutron energy transfer $\approx3.4$~meV. 
Given that the feature becomes more pronounced on cooling, 
this $T$-dependence suggests that it is likely associated with 
spin-wave formation due to long-range magnetic order. 
The difference between the $T = 1.6$~K and $T = 20$~K spectra 
[Fig.~\ref{fig7}(d)] reveals the magnetic spectrum more clearly, 
showing a band of excitations that exists below 3.3~meV. 
Below 1~meV, the high temperature background is large, 
leading to an over-subtraction of the data. 
The energy scale, wave-vector and temperature dependence of the scattering below 
1~meV indicate that the large background may be attributed to acoustic 
phonons that contribute to the scattering intensity at these energies. 
Furthermore, a small percentage of hydrogen in the sample due to 
incomplete deuteration could increase the background due to the 
large incoherent cross-section of the $^1$H isotope. 
For $T < T_{\rm c}$, magnetic Bragg peaks along 
$E = 0$ are evident at $|{\bf Q}| \approx 0.8$~\AA$^{-1}$, 
1.6~\AA$^{-1}$ and 2.0~\AA$^{-1}$
[white arrows in Fig. 7(d)] which is fully consistent with the elastic 
neutron diffraction patterns. 

\begin{figure}[t]
\centering
\includegraphics[width=8.5cm]{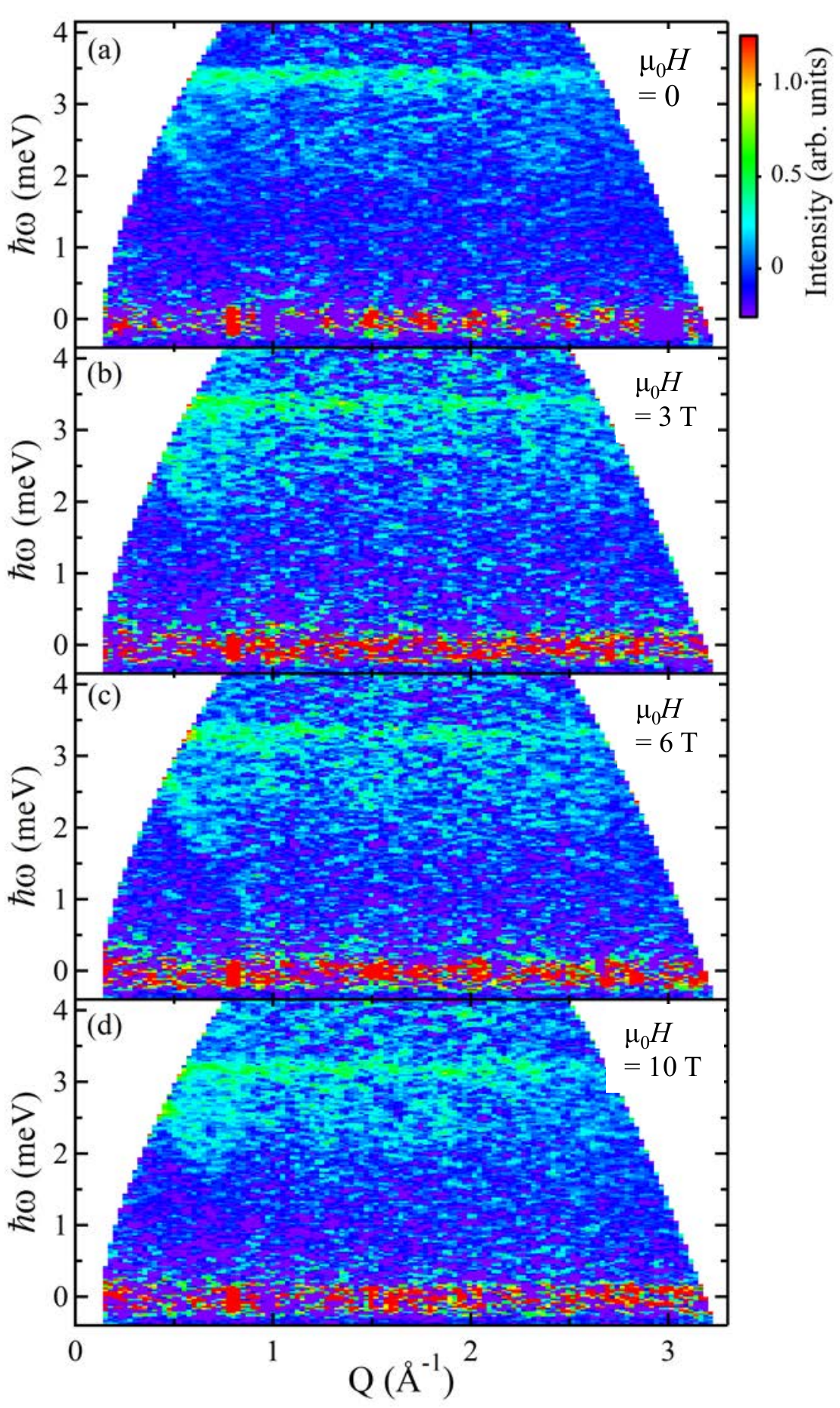}
\caption{Background-subtracted magnetic-field dependent INS data 
obtained at $T=1.5$~K for (a)~$\mu_0 H = 0$, (b)~$\mu_0 H = 3$~T, 
(c)~$\mu_0 H = 6$~T, and (d)~$\mu_0 H = 10$~T. 
Data were binned and smoothed in a similar manner to those in Fig.~\ref{fig7}. 
Data measured at $T = 20$~K and at corresponding magnetic fields were
used as a paramagnetic backgrounds for each set of data taken at 1.5~K.
}
\label{fig8}
\end{figure}

\begin{figure}[t]
\centering
\includegraphics[width=8.5cm]{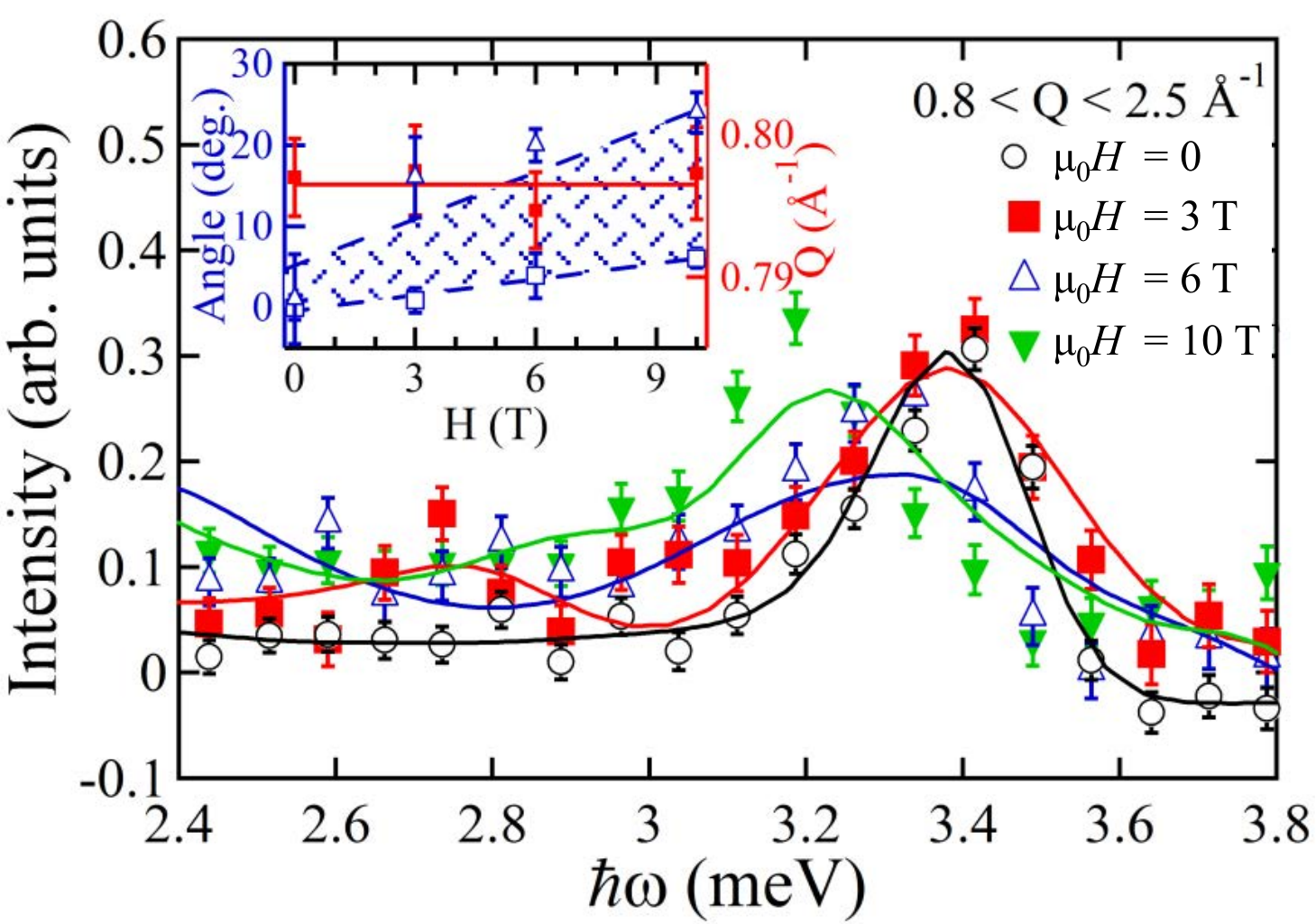}
\sloppypar
\caption{(Main plot): Background subtracted intensity as a function of 
energy transfer $(\hbar \omega)$ for magnetic fields $\mu_0H \leq 10$~T. 
The data correspond to those shown in Fig.~\ref{fig8} integrated between 0.8 and 
2.5~\AA$^{-1}$ in wave-vector transfer. The solid lines correspond to the 
spin-wave calculations discussed in the text; they use a randomly applied magnetic 
field orientation. (Inset)~The open blue triangles correspond to the 
field-dependent spin reorientations of the Ni(II) moments using the 
random-field calculation.  The open blue squares correspond to the 
field-dependent spin reorientations of the Ni(II) moments considering only 
moments applied along the $c$-axis of the crystal.  
The shaded area is the range between a linear field-dependent fit to this 
reorientation angle for each of the models.  The solid red squares~(right axis) 
in the inset show the location of the magnetic Bragg peak as a 
function of  $H$.  The solid red line is a fit to a constant value.
}
\label{fig9}
\end{figure}

Figure~\ref{fig8} shows the magnetic-field dependence of the excitation 
spectrum. For each data set, a $T = 20$~K background measurement at the 
same applied magnetic field was made and subtracted from subsequent 
field-dependent data. We found these $20$~K spectra to give a reasonable
representation of the behavior 
of the paramagnetic phase of our material. 
The peak in the zero-field spin-wave density-of-states shifts to lower 
energy transfers as the magnetic field increases. 
The magnetic Bragg peak at $Q = 0.8$~\AA$^{-1}$
does not shift with increasing magnetic field, indicating that the periodicity of 
the long-range order does not change with field (see inset of Fig.~\ref{fig9}). 
For non-zero magnetic fields, [Fig.~\ref{fig8}(b-d)], there is some additional 
scattering intensity that is present at energies higher 
than the top of the zero-field spin wave band at 3.4~meV.
 
Integrating the data from Fig.~\ref{fig8} for momentum transfers in the range 
$0.8 -2.5$~\AA$^{-1}$, (Fig.~\ref{fig9}) 
illustrates how the intense portion of the zero-field spin-wave mode at an 
energy transfer of 3.4~meV softens with the applied magnetic field. 
There is also an increase in scattering intensity at higher energy 
transfers with increasing field indicating that the degeneracy of the 
spin-wave mode is lifted by the field and a portion of the spectrum 
is moving to larger energy transfers.

With the material well within the ordered magnetic phase at $T = 1.6$~K,
we modeled the zero-field spectrum shown in Fig.~\ref{fig7}(d) using the 
numerical spin wave calculation package SPINW~\cite{a35}.
We use the aforementioned zero-field magnetic structure as the basis 
for all subsequent simulations. The exchange interactions $J$ and $J'$ 
are included in the model as well as the anisotropy term $D$ as described in 
Eqn.~\ref{eqn1}.. A powder-average spin-wave spectrum is calculated 
for energy transfers between 2 and 4~meV for a large range of values in 
$D$, $J$ and $J’$ parameter space. The limited range of energy transfer was 
used for comparison to avoid the lower energy region due to the acoustic-phonon 
scattering previously described. 
Each simulated spin-wave spectrum was directly compared to the measurement 
and a value of reduced $\chi^2$ was determined for each triplet of energy 
parameters. Fitting parameters for each simulation included a constant 
background and an overall multiplicative prefactor to scale 
the calculated scattering intensity~\cite{a57,a58}. 
The best-fit parameters were found from the minimum in reduced $\chi^2$. 
Figure~\ref{fig7}(e) shows the final simulated spectrum. 
The resulting terms in the Hamiltonian were determined to be $D = 13.3(1)$~K, 
$J = 10.4(3)$~K and $J' = 1.4(2)$~K, with $D$ and $J$ falling 
in the range estimated from thermodynamic measurements. 
Any potential next-next-nearest neighbor Ni---Ni exchange interaction along 
the $a-$ and $b-$axes ({\it i.e.}, the diagonals of the [Ni(pyz)$_2$]$^{2+}$ 
square plaquettes, see Fig.~\ref{fig2}), with a distance of 9.880~\AA, 
would likely be much smaller than $J'$ and require measurements on 
single crystals with improved energy resolution to determine accurately. 

\begin{figure}[t]
\centering
\includegraphics[width=8.5cm]{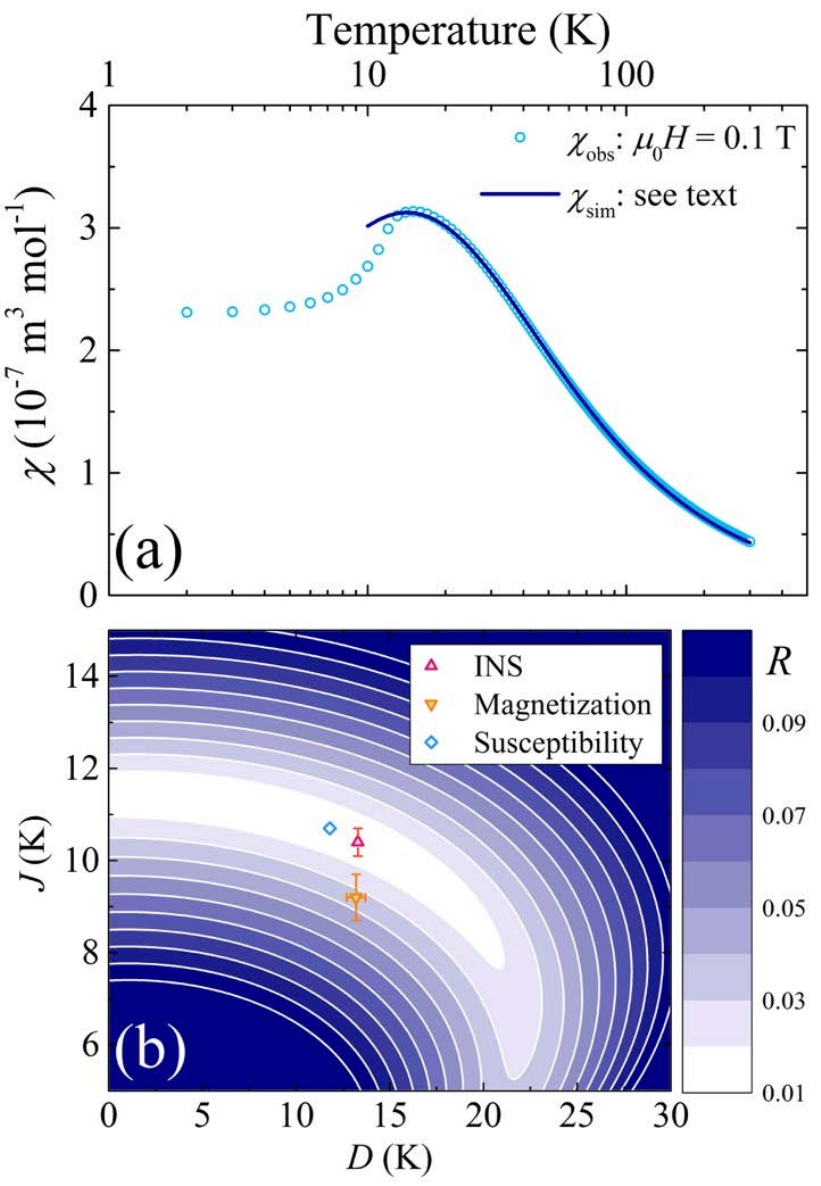}
\sloppypar
\caption{(a)~Magnetic susceptibility $\chi$ versus temperature $T$ data for 
[Ni(HF$_2$)(pyz)$_2$]SbF$_6$ $(\mu_0 H = 0.1$~T) interpolated to evenly 
spaced temperatures in the range $2 \leq T \leq 300$~K, $\Delta T = 1$~K 
($\chi_{\rm obs}$, points). (b)~Contour plot of the quality factor $R$ in the 
$D,J$ plane found by comparing the simulated susceptibility 
(Eqns.~\ref{eqn6} and \ref{eqn7}) to the measured data using Eqn.~\ref{eqn8}
(see text for details). The result shows bands of $D$ and $J$ 
that provide very similar quality fits to the data. 
The model $(\chi_{\rm sim})$ that yields the minimum value of $R$ 
(diamond) is displayed as a solid line in panel (a).
}
\label{fig10}
\end{figure}

In addition, we calculated the magnetic-field dependence of the 
excitation spectrum. For this case, we perform the powder average using 
a random field direction with the moments tilted by a fixed angle from their 
zero-field orientation toward the applied magnetic field.  
We then average this calculated spectrum over 128 random 
applied field directions.  To achieve the correct weighting, each 
calculated spectrum was normalized by a factor corresponding to the 
cosine of the angle between the crystallographic $ab$-plane and the direction 
of the applied magnetic field.  For each fixed magnitude of applied magnetic 
field, we use the zero-field determined exchange constants and vary the 
magnitude of the spin-canting angle to compare the calculated spectrum 
to the measured INS spectrum, using an additive background and overall 
multiplicative prefactor. The field-dependence of the spin-canting angle 
determined from this method is shown in the inset of Fig.~\ref{fig9}. 
Here, the error bars correspond to an increase in reduced $\chi^2$ by 2.5\%. 
The resulting lineshapes drawn in Fig.~\ref{fig9} are in good agreement 
with the measured field-dependent spectra. Because of the anisotropy 
intrinsic to our sample, it is possible that a number of the grains of the powder 
would be rearranged by the applied magnetic field.  
To understand this possibility, we considered the spectrum in Fig.~\ref{fig9} 
as if the magnetic field was solely applied along the $c$-axis of the compound. 
From examination of the spin-wave scattering intensity, 
the $c$-axis component of magnetic field is mostly responsible for the 
shift in the density of magnetic states to lower energy transfers. 
In this case, the spin canting angle determined as a function of applied magnetic 
field is smaller than the random-field calculation, as is shown in the 
inset of Fig.~\ref{fig9}. For both calculations, we find that the ordered 
magnetic moments gradually rotate from their zero-field orientation toward the 
applied magnetic field as a function of the applied magnetic field. 
The powder nature of the sample does not allow us to easily model 
the case of a distribution of ordered moment orientations. 
Considering a linear dependence for the ordered moment direction 
as a function of field for both models, the shaded range in the 
inset of Fig.~\ref{fig9} represents the most likely range of canting angles 
for different orientations of applied magnetic field. 

\begin{figure}[htbp]
\centering
\includegraphics[width=8.5cm]{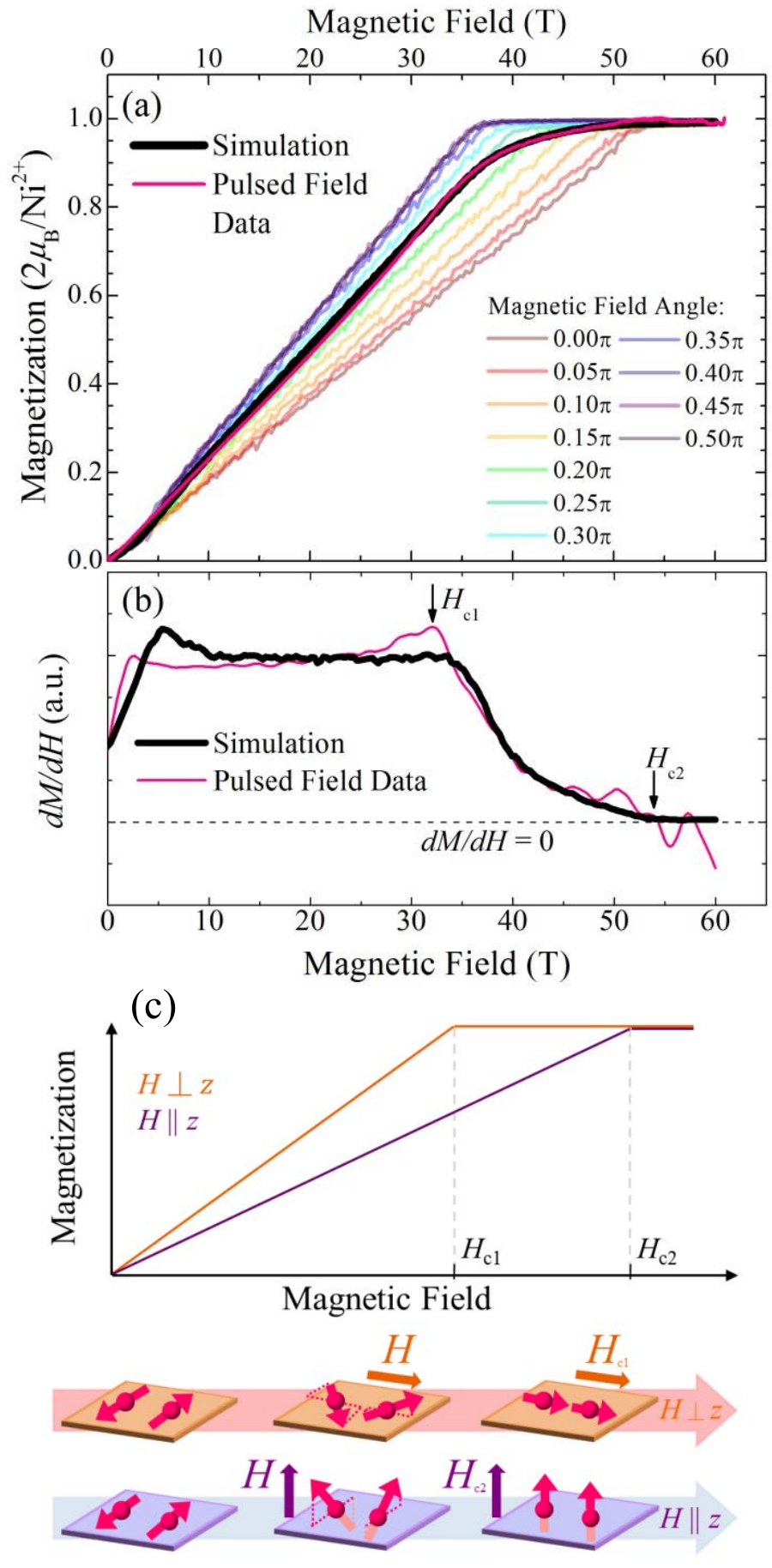}
\caption{(a)~Simulated magnetization $(M)$ versus magnetic field for 
[Ni(HF$_2$)(pyz)$_2$]SbF$_6$ using a Monte-Carlo energy-minimization routine 
for an 8-spin cluster governed by Eqn.~\ref{eqn1} with $D = 13.3$~K, $J = 10.3$~K, 
$J' = 1.43$~K and $g = 2.08$. Magnetization curves are obtained 
for 21 orientations of the magnetic field with respect to the hard 
axis (colored lines show the unique angles) and the powder-average $M$ 
(black line) is determined from Eqn~\ref{eqn2}. Good agreement with the 
$T = 0.56$~K pulsed-field $M$ data (pink line) is achieved. 
(b)~Differential susceptibility of [Ni(HF$_2$)(pyz)$_2$]SbF$_6$ 
as deduced from the Monte-Carlo simulation (black line) compared with 
pulsed-field measurements (pink line). 
(c)~Illustration of the field-induced spin reorientation that occurs 
when $H$ is applied perpendicular or parallel to the $z$-axis. 
}
\label{fig11}
\end{figure}

\vspace{3mm}
\noindent
{\it Magnetic susceptibility} 
The magnetic susceptibility $(\chi_{\rm obs})$ of 
[Ni(HF$_2$)(pyz)$_2$]SbF$_6$ exhibits a 
broad maximum at 16~K associated with the formation of spin correlations 
along the Ni---FHF---Ni chains as the sample is 
cooled~\cite{a18} [Fig.~\ref{fig10}(a)].
There is a rapid decrease in $\chi_{\rm obs}$ on reducing 
$T$ through the ordering 
transition at 12.2~K and the susceptibility plateaus 
as $T$ is 
decreased further.
The data for $T \geq 10$~K are compared to a simulation of the 
susceptibility $(\chi_{\rm sim})$ using a 1D chain model~\cite{a59} 
of $S = 1$ ions with intrachain exchange $J$ and single-ion anisotropy $D$, 
which is expressed as:
\begin{equation}
\chi_{\rm sim} = \frac{2 \mu_0 N_{\rm A} \mu_{\rm B}^2g^2}{3k_{\rm B}J}
\left[\frac{t^2+0.5t +0.1}{a_3t^3+a_2t^2+a_1t+a_0}\right]
\label{eqn6}
\end{equation}
where $t=\frac{T}{J}$ is the reduced temperature and the coefficients
$a_i$ are polynomials of the $\frac{D}{J}$ ratio:
\begin{equation}
a_i=\sum_{j=0}^2c_{ij}\left(\frac{D}{J}\right)^j
\label{eqn7}
\end{equation}
with the coefficients $c_{ij}$  given in Table~\ref{table7}. 
Eqn.~\ref{eqn6} can be used to attempt fits of $\chi(T)$ data for powders.

Fixing the powder average $g$-factor to the published value~\cite{a18}, 
$g = 2.08$, the susceptibility (Eqn.~\ref{eqn6}) was simulated for 
parameters in the range $0 \leq D \leq 30$~K, $5\leq J \leq 15$~K 
and $10 \leq T \leq 300$~K ($\Delta T = 1$~K). 
The quality factor
\begin{equation}
R=\frac{\sum_i|\chi_{{\rm obs},i}-\chi_{{\rm sim},i}|}{\sum_i\chi_{{\rm obs},i}}
\label{eqn8}
\end{equation}
was computed for each simulated curve, where $i$ runs over all data 
points for $T \geq 10$~K. A contour plot of $R$ across the $D-J$ plane 
[Fig.~\ref{fig10}(b)] reveals bands of $D$ and $J$ values that all produce
$\chi_{\rm sim}$ curves providing equally good representations of the 
measured data. This insensitivity of Eqn~\ref{eqn6} to a detuning of $D$ and $J$ 
from the minimum in $R$ (white region) indicates that the results of fitting 
the susceptibility to Eqn~\ref{eqn6} should be treated with caution. 
The parameters $D$ and $J$ are strongly correlated, which likely results 
from their competing energy scale and the fact that a low-field bulk 
measurement of the susceptibility, as obtained from a powder sample, cannot 
differentiate between the effects of spatial exchange anisotropy 
and single-ion anisotropy for an $S = 1$ chain.

\begin{table}[t]
\centering
\caption{Coefficients ($c_{ij}$, Eqn.~\ref{eqn7}) for the 
polynomials ($a_i$, Eqn.~\ref{eqn6}) from Ref.~\onlinecite{a59}. 
}
\label{table7}
\begin{tabular}{lrrr}
\hline\hline
Polynomial ($a_i$) & $c_{i0}$   & $c_{i1}$     & $c_{i2}$ \\
\hline
 $i=0$                  & 1.67268034    & -0.26449121  & -0.102945  \\
 $i=1$                 & 1.710151691  & 0.5114739    & 0.18853874 \\
 $i=2$                  & 1.899474528  & -0.166396406 &  0.1494167 \\
 $i=3$                  & 1          & 0            & 0            \\
\hline\hline
\end{tabular}
\end{table}

The values of $D$ and $J$ deduced from the mean-field analysis of 
the magnetization and the more precise results obtained by simulating 
the INS data (Fig.~\ref{fig10}(b), triangles) both fall close to the white 
band in the contour map, indicating good consistency with the broad 
temperature dependence of the $\chi_{\rm obs}$ data. 
The high-field magnetization and 
INS measurements, however, offer a significant advantage over the 
susceptibility analysis, as a single experiment can constrain both $D$ and $J$.
This results from two key differences in the techniques as compared 
to the low-field susceptibility measurements: 
(i)~the high-field magnetization is sensitive to the anisotropy of the critical field; 
and (ii)~a local probe such as neutron scattering is sensitive to the 
local symmetry of the magnetic centers and so a single experiment
can constrain both $D$ and $J$. 
Thus, the choice of a high-field or local experimental probe was crucial to 
differentiate the effects of spatial-exchange and single-ion 
anisotropy in the magnetic properties of this sample. 
Furthermore, the simulated susceptibility curve that 
minimizes $R$ [Fig.~\ref{fig10}(a), line] significantly deviates 
from the measured data in the ordered phase. 
For temperatures $T \leq T_{\rm c}$, the effects of finite interchain 
interactions $(J')$ may not be ignored and this further complicates 
the analysis of the susceptibility at low temperatures. 
Thus, we find that the local INS probe is necessary to efficiently and precisely 
determine the full set of magneto-structural parameters $(D, J$ and $J')$ 
for [Ni(HF$_2$)(pyz)$_2$]SbF$_6$.

\begin{figure}[htbp]
\centering
\includegraphics[width=8cm]{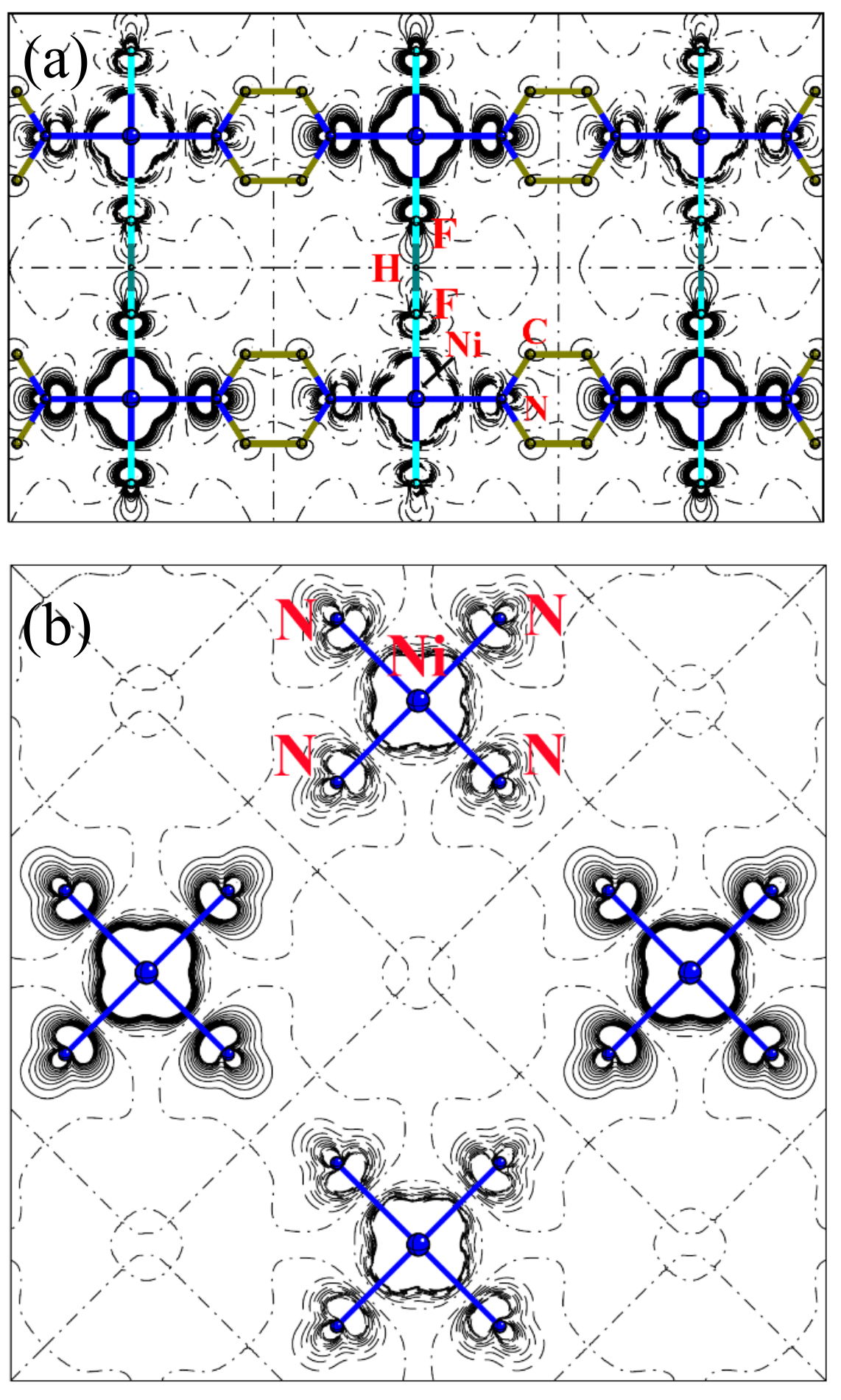}
\caption{The electron spin-density distribution in the (a)~$[110]$ and 
(b)~$[001]$ planes, calculated for the AFM state (solid and dashed 
lines represent the excess or defect of $\alpha$ spin-density; 
contours are drawn with a logarithmic increase). 
The spin density is delocalized along both ligand types. However,
 the spin delocalization is essentially quenched along Ni---pyz---Ni 
because only the $\sigma$ skeleton is involved. 
In contrast, the spin delocalization is not disrupted 
along the Ni---FHF---Ni bridge. 
}
\label{fig12}
\end{figure}

\subsection{Calculations and modeling}

\noindent
{\it Monte-Carlo simulation of the magnetization.}
The magnetization of [Ni(HF$_2$)(pyz)$_2$]SbF$_6$ was calculated from a 
Monte-Carlo energy minimization routine on an 8-spin cluster for 21 orientations 
of the applied field with respect to the magnetic hard-axis of the Ni(II) ions
[Fig.~\ref{fig11}(a), 
colored lines]. The critical field was found to be anisotropic, such that 
a greater applied field is required to saturate the moments as the field 
orientation is moved away from the easy-plane. The resultant powder-average 
magnetization was determined from Eqn.~\ref{eqn2} (thick line) and compared 
to the measured magnetization at the lowest available temperature 
(pink line; $T = 0.56$~K). An overall good agreement of the rounded 
approach towards saturation for the powdered sample is obtained. 
Furthermore, the parameters $D = 13.3$~K, $J = 10.3$~K, $J' = 1.43$~K 
and $g = 2.11$ yield a good quantitative agreement for the 
two observed critical fields [Fig.~\ref{fig11} (a)].

The measured pulsed-field magnetization develops a slight concavity on 
approaching $H_{\rm c1}$ with increasing field as the bath 
temperature is reduced below 8~K [Fig.~\ref{fig11}(b)], which leads to a 
small rise in ${\rm d}M/{\rm d}H$ at $H_{\rm c1}$. 
This behavior is not reproduced by the simulation that 
employed classical vectors to represent the Ni(II) moments. 
The small discrepancy may be attributed to the development of 
quantum fluctuations of the $S = 1$ magnetic moments, which 
result from the Q1D nature of the spin-exchange interactions~\cite{a60} 
that act to suppress the magnetization at low temperatures. 
Furthermore, the small rise in ${\rm d}M/{\rm d}H$ evident close to 
5~T in the simulated data is an artefact attributed to the finite number 
of cycles used to determine the ground state spin configuration. 
This ultimately causes a slight underestimation of the 
magnetization in this field regime~\cite{a61}.

\begin{table}[]
\centering
\caption{Magnetic parameters from Eqn.~\ref{eqn1} (in Kelvin) as 
deduced from high-field magnetization data and inelastic neutron-scattering (INS) 
experiments. For comparison, DFT-computed values are included in the 
last column. To decompose $n\langle J\rangle$ into individual $J$ and $J'$ 
contributions from the magnetization data it was necessary to infer an 
average value of $J'$ from a catalog of related coordination polymers 
that also contain square [Ni(pyz)$_2$]$^{2+}$ motifs (see Table~\ref{table6}).
}
\label{table8}
\begin{tabular}{lrrr}
\hline\hline
Parameter (K) & $M(H)$ data & INS data & DFT \\
\hline
$D$             & 13.2(5)   & 13.3(1)  & --   \\
$J$             & 9.2(5)    & 10.4(3)  & 9.2 \\
$J'$            & 0.9(2)    & 1.4(2)   & 1.8 \\
\hline\hline
\end{tabular}
\end{table}

Fig.~\ref{fig11}(c) shows a schematic of the behavior expected for 
classical spins as the field is applied parallel and perpendicular to 
the hard axis, with anisotropic critical fields indicated. 
In real Q1D $S = 1$ systems the magnetization at fields less than $H_{\rm c}$ 
will not be a linear function of field; instead the reduced dimensionality gives 
rise to a concavity in the single-crystal $M(H)$ data, 
as captured by quantum Monte Carlo simulations~\cite{a60}.

\vspace{3mm}
\noindent
{\it Theoretical spin-density distribution.}
Periodic DFT calculations on [Ni(HF$_2$)(pyz)$_2$]SbF$_6$ enabled us to 
estimate the $J$ and $J'$ exchange constants by calculating the energy 
of the FM state and of two different AFM states, with spin pairing 
along the $c$-axis (AFM$_{\rm FHF}$) or in the 
$ab$-plane (AFM$_{\rm pyz}$), as well as the full AFM 
state with both kinds of pairing (Fig. S2). The energy difference $\Delta E$ 
between the FM and the AFM$_{\rm FHF}$ or AFM$_{\rm pyz}$ 
states can be used to 
calculate $J$ or  $J'$, respectively, using Eqn.~\ref{eqn1} and assuming the 
calculated $\Delta E$ to be equivalent to the result of the corresponding 
Heisenberg Hamiltonian. The primary $J$ can also be obtained from 
$E({\rm AFM}) - E({\rm AFM_{pyz}})$, whereas $J'$ can be obtained from 
$E({\rm AFM}) - E({\rm AFM_{FHF}})$. Considering the experimental 
geometries, both approaches provided $J = 9.2$~K and $J' = 1.8$~K, 
in good agreement with the model derived from INS. 
The calculated periodic wave function also enabled mapping of the 
spin-density distribution in all states. 
Fig.~\ref{fig12} shows the spin-density distribution of the AFM state. 
For all states (FM or AFM), the Ni atom bears $\approx 1.75e$ 
of excess spin. The remaining $0.25e$ is delocalized onto all ligands 
and are responsible for the observed magnetic exchange. 
Despite $ J' \ll J$, the largest spin population lies on the pyrazine 
N-atom $(\approx 0.05e)$, whereas only $0.02e$ reside on the F-atoms. 
This is certainly caused by N being a stronger donor than F$^-$ 
(in accord with the spectrochemical series). 
However, because the exchange mechanism through pyrazine is 
mainly $\sigma$-type (as shown for some 
Cu-pyrazine networks)~\cite{a62} it is not as effective for the magnetic 
exchange, which explains the smaller $J'$. The optimal delocalization via 
the pyrazine-bridge would occur through the $\pi-$electrons which are not 
significantly involved in the spin density. 
In fact, population analysis shows that the atomic $p$ orbitals of N and C involved 
in the pyrazine $\pi-$system contribute very little to the overall spin
density. On the other hand, the short F---F distance in the HF$_2^-$ ligand likely 
promotes more effective spin delocalization, as clearly evident in Fig.~\ref{fig12}(a).
 
\begin{figure}[h]
\centering
\includegraphics[width=8.5cm]{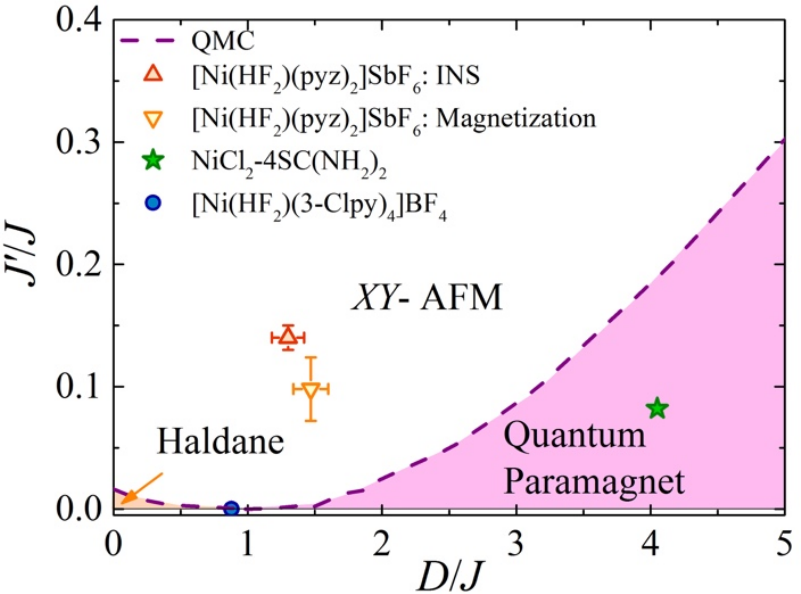}
\sloppypar
\caption{ Phase diagram of $S = 1$ Q1D materials as a function of intrachain 
exchange $(J)$, interchain interaction $(J')$ and single-ion anisotropy $(D)$. 
The phase boundaries are the results of quantum Monte-Carlo calculations~\cite{a9}.
The position of [Ni(HF$_2$)(pyz)$_2$]SbF$_6$ as determined from both 
pulsed-field magnetization and INS (triangles) predicts the material 
to have an antiferromagnetic $XY$-ordered ground state. 
The positions of the quantum paramagnet~\cite{a4} NiCl$_2$-4SC(NH$_2$)$_2$ (star) 
and Haldane-chain system~\cite{a8} [Ni(HF$_2$)(3-Clpy)$_4$]BF$_4$ 
(circle) are included for comparison.
}
\label{fig13}
\end{figure}

\section{Conclusions}
The material [Ni(HF$_2$)(pyz)$_2$]SbF$_6$ retains tetragonal symmetry 
for temperatures down to 1.5~K and may be characterized as a Q1D $S = 1$ 
quantum magnet that exhibits a 3D $XY$-AFM ground state below 12.2(1)~K 
as determined from high-resolution elastic neutron scattering. The magnetic 
properties can be described by the Hamiltonian in Eqn.~\ref{eqn1}, 
with $D = 13.3(1)$~K, $J = 10.3(3)$~K 
and $J' = 1.4(2)$~K as determined by INS measurements. 
We showed that these values are in reasonable agreement with the 
initial estimates deduced from high-field magnetization studies (see Table~\ref{table8}), 
while low-field bulk thermodynamic probes such as magnetic susceptibility 
were unable to satisfactorily untangle the effects of the spatial exchange and 
single-ion anisotropies. Compared to the previous study~\cite{a18}, 
the DFT results presented in this work more closely agree with the 
experimentally derived $J$ and $J'$ parameters.
 
The predicted~\cite{a9} phase diagram for easy-plane $(D > 0)$ 
Q1D spin-1 systems is shown in Fig.~\ref{fig13}. 
The phase boundaries, as deduced from QMC calculations, 
separate regions of $XY$–AFM order from the disordered quantum 
paramagnetic and Haldane phases; the three phases converge~\cite{a9} 
at the quantum critical point $\frac{D}{J}= 0.97$. 
The relative position of [Ni(HF$_2$)(pyz)$_2$]SbF$_6$ is indicated on 
the phase diagram both the precise $J'/J$ and $D/J$ ratios from INS 
and those estimated from a mean-field analysis of the 
critical fields observed in pulsed-field magnetization. 
Both estimates predict [Ni(HF$_2$)(pyz)$_2$]SbF$_6$ to exhibit an $XY$-AFM 
ground state, though only INS independently determined all three 
parameters needed to validate this prediction. 
The lack of a gapped ground state is in full agreement with both the 
field-dependent heat capacity, which showed clear evidence of a 
transition to an AFM ground state, and the refined magnetic 
structure from elastic neutron scattering in zero field.

Given the proximity of [Ni(HF$_2$)(pyz)$_2$]SbF$_6$ to the 
quantum-critical point and phase boundaries to two other 
distinct phases, this material is an intriguing candidate for a 
pressure-dependent study to determine the extent to which the 
parameters $D$, $J$ and $J'$ can be tuned to explore the phase 
diagram and further test the predictions of QMC 
simulations for Q1D $S = 1$ systems. While there is no substitute 
for the detailed study of single crystals, the success of this 
work on powder samples demonstrates the complementary capabilities 
of the micro- and macroscopic probes involved. 
More specifically, the experimental sequence was as follows: (i)~micro-crystal 
synchrotron X-ray diffraction determines the crystal structure; (ii)~using this 
structure as a starting point, an analysis of the powder elastic-neutron diffraction 
establishes the magnetic ground state and charts the evolution of the 
ordered moment as a function of temperature; (iii)~using 
the magnetic structure as the starting point, an analysis of the powder inelastic-neutron 
scattering determines the magnetic exchange and anisotropy 
parameters; (iv)~an independent estimate of the magnetic parameters was 
possible via a careful analysis of the high-field powder magnetometry data 
and was found to be in good agreement with the results of the 
neutron scattering; (v)~the field-temperature phase diagram was mapped out 
using high-field-magnetization and heat-capacity measurements. 
Having established the applicability of the methodology, we are currently 
applying this experimental protocol to other $S = 1$ materials.

\section{Acknowledgments}
The work at EWU was supported by the National Science Foundation (NSF) 
under grant no. DMR-1306158 and by the National Institute of Standards 
and Technology (NIST) Cooperative Agreement 70NANB15H262. 
ChemMatCARS Sector 15 is principally supported by the 
Divisions of Chemistry (CHE) and Materials Research (DMR), NSF, 
under grant no. CHE-1346572. Use of the Advanced Photon Source, 
an Office of Science User Facility operated for the U.S. Department of 
Energy (DoE) Office of Science by Argonne National Laboratory, 
was supported by the U.S. DoE under Contract No. DE-AC02-06CH11357. 
We acknowledge the support of NIST, U.S. Department of 
Commerce (DoC), in providing their neutron research facilities 
used in this work; identification of any commercial product 
or trade name does not imply endorsement or recommendation 
by NIST. Work performed at the National High Magnetic Field 
Laboratory, USA, was supported by NSF Cooperative Agreement 
DMR-1157490, the State of Florida, U.S. DoE, and through 
the DoE Basic Energy Science Field Work Project “Science in 100 T.” 
M.B.S. was supported by the Scientific User Facilities Division, 
Office of Basic Energy Sciences, U.S. Department of Energy. 
We gratefully acknowledge the ISIS-RAL facility for the provision 
of beamtime. J.B. thanks EPSRC for financial support. P.A.G. 
acknowledges that this project has received funding from the 
European Research Council (ERC) under the European 
Union's Horizon 2020 research and innovation programme 
(grant agreement No. 681260). Research at the University 
of Bern was funded by the Swiss NSF (project 160157).
J.S. acknowledges a Visiting Professorship from
the University of Oxford that enabled some of the 
experiments reported in this paper. 

\vspace{5mm}

\noindent
$^{\ddag}$Contact email: J.D.Brambleby@warwick.ac.uk\\
$^*$Contact email:  jmanson@ewu.edu\\
$^\aleph$Contact email: jsingle@lanl.gov\\


\begin{thebibliography}{99}
\bibitem{HCl}
Caution! Hydrofluoric acid must be handled with 
great care as it is highly toxic, corrosive, and will cause 
severe burns. Proper personal protective equipment must 
be worn at all times regardless of the amount of HF being used.
\bibitem{nonneel}
We refer to the ordering temperature as $T_{\rm c}$,
rather than $T_{\rm N}$, as our system is not
a conventional N\'{e}el antiferromagnet.
\bibitem{a61}
At low fields, where the number of degenerate ground states in the 
uniaxial system is greatest, running a finite length Monte-Carlo 
Markov chain simulation to iterate an ensemble 
of spins towards their ground state configuration can lead 
to a small proportion of the spins ending up perturbed 
from their lowest energy orientation. Through averaging 
the final magnetization over many repeated simulations 
performed in this manner, this can lead to an underestimation 
of the magnetization and the slight kink in $M$ versus $H$ 
as found in the simulated traces.
\end{thebibliography}
\end{document}